%% file: SemiEfarxiv.tex
\documentclass[11pt,a4paper,reqno]{amsart}
\usepackage{multirow}
\usepackage[centertags]{amsmath}
\usepackage{amsfonts}
\usepackage{amssymb}
\usepackage{amsthm}
\usepackage{newlfont}
\usepackage{a4}
\usepackage{enumerate}
\usepackage[pdftex]{graphicx,color}
\theoremstyle{definition}

\theoremstyle{remark}
\newtheorem{example}{Example}[section]
\usepackage{bbm}
\newcommand{\ii}{\mathrm{i}}
\newcommand{\map}{\rightarrow}
\newcommand{\intt}{\operatorname{int}}

\newcommand{\G}{{\cal G}}

\newcommand{\q}{\quad}

\renewcommand{\epsilon}{\varepsilon}

\newcommand{\ep}{\varepsilon}
\newcommand{\la}{\lambda}
\newcommand{\al}{\alpha}
\newcommand{\om}{\omega}
\renewcommand{\rho}{\varrho}
\renewcommand{\phi}{\varphi}

\newcommand{\R}{{\mathbb{R}}}

\newcommand{\N}{{\mathbb N}}

\newcommand{\Z}{\mathbb{Z}}
\newcommand{\C}{\circ}

\newcommand{\set}[2]{\left\{#1 \, |\, #2 \right\}}
\newcommand{\setb}[2]{\left\{#1 \, \mid\, #2 \right\}}

\newcommand{\abs}[1]{\left\vert#1\right\vert}

\newcommand{\sca}[2]{\langle #1,\, #2\rangle}

\def\R{\mathbb R}
\def\Z{\mathbb Z}
\def\N{\mathbb N}
\def\C{\mathbb C}

\def\O{\mathop{\operatorname{{O}}}\nolimits}

\def\SU{\mathop{\operatorname{{SU}}}\nolimits}

\def\G{\mathop{\operatorname{{G}}}\nolimits}

\def\Aone{\mathop{\operatorname{{A_1}}}\nolimits}
\def\Atwo{\mathop{\operatorname{{A_2}}}\nolimits}

\def\Ctwo{\mathop{\operatorname{{C_2}}}\nolimits}
\def\Gtwo{\mathop{\operatorname{{G_2}}}\nolimits}

\begin{document}
\title[Semisimple $E-$functions]
{On $E-$functions of Semisimple Lie Groups}
\author{Ji\v{r}\'{\i} Hrivn\'{a}k$^{1,2}$}
\author{Iryna Kashuba$^{3}$}
\author{Ji\v{r}\'{\i} Patera$^1$}

\date{\today}
\begin{abstract}\
We develop and describe continuous and discrete transforms of
class functions on a compact semisimple, but not simple, Lie group
$G$ as their expansions into series of special functions that are
invariant under the action of the even subgroup of the Weyl group of $G$. We
distinguish two cases of even Weyl groups -- one is the direct
product of even Weyl groups of simple components of $G$, the
second is the full even Weyl group of $G$. The problem is rather
simple in two dimensions. It is much richer in dimensions greater
than two -- we describe in detail $E-$transforms of semisimple
Lie groups of rank $3$.
\end{abstract}\

\maketitle

\noindent
$^1$ Centre de Recherches Math\'ematiques,
         Universit\'e de Montr\'eal,
         C.~P.~6128 -- Centre Ville,
         Montr\'eal, H3C\,3J7, Qu\'ebec, Canada; patera@crm.umontreal.ca\\
$^2$ Department of Physics, Faculty of Nuclear Sciences and
Physical Engineering, Czech Technical University in Prague,
B\v{r}ehov\'a~7, 115 19 Prague 1, Czech Republic;
jiri.hrivnak@fjfi.cvut.cz\\ $^3$ Departamento de Matem\'atica,
Instituto de Matem\'atica e Estat\'istica, Universidade de S\~ao
Paulo, Rua do Mat\~ao 1010, CEP 05508-090, S\~ao Paulo, Brasil;
kashuba@ime.usp.br


\section{Introduction}\label{IntroSection}

The paper can be considered as a completion of \cite{PK} and also
\cite{KP1} by the most complicated and interesting and potentially
most useful version of the problem of the $E-$transform and their
special functions ($E-$functions) of any semisimple, but not
simple, compact Lie group $G$. The rank $n$ of $G$ is the number
of real variables of the $E-$functions.

Assuming that $G=G_1\times G_2$, there are two possibilities to
introduce, what is then called, the even subgroup $W^e$ of the
Weyl group $W$ of $G$.  Both options are mentioned in
\cite{PK,KP1} but only the simpler one of the two is considered
there.  Both options have the most valuable property: they admit
development of multidimensional Fourier transforms, although
rather different in each case. It is natural to expect that of the
two versions each will find its field of optimal applications.
Therefore our aim in here is to consider the option that was
passed over in \cite{PK,KP1}.

Suppose a compact semisimple Lie group $G$ is a product of two
simple invariant subgroups, $G=G_1\times G_2$. We are interested
in the even subgroup $W^e$ of the Weyl group of $G=G_1\times G_2$.
The simpler of the two options is to take as $W^e$ the product
$W^e(G_1)\times W^e(G_2)$, while the second option, investigated
here, is to consider the full even subgroup $W^e(G_1\times G_2)$.
Since
\begin{gather*}
W^e(G_1\times G_2)\supset W^e(G_1)\times W^e(G_2)\,,
\end{gather*}
the second option is clearly richer. There is an important
implication of this fact for the class of functions that can be
expanded in each option.

For example, in the simplest case when $G_1$ and $G_2$ are rank 1
simple Lie groups, $W^e(G_1)\times W^e(G_2)$ is of order 1, while
$W^e(G_1\times G_2)$ is of order 2.

The rank 3 groups are considered all in detail in view of their
likely applicability. The $E-$functions of none of these cases were
studied before with the exception $G=A_1\times A_1\times A_1$
which is a straightforward concatenation of three 1-dimensional
cases oriented in mutually orthogonal directions.

The pertinent standard properties of simple Lie groups and their
$E-$functions are collected in section~\ref{general}. Two types of
even Weyl groups for semisimple Lie groups are introduced in
section~\ref{EvenWeyl}. The continuous and discrete orthogonality
of corresponding $E-$functions is also included there. Explicit
formulas of $E-$transforms for semisimple Lie groups of rank less
or equal to three are contained in section~\ref{Etrans}. Comments
and follow-up questions are listed in section~\ref{Conc}.
\section{$E-$functions and their properties}
\label{general}

In this section we collect some facts from the basic theory of
simple Lie groups, the Weyl groups and their orbit functions. More
details can be found in \cite{Hum} (for Lie groups), \cite{Kane},
\cite{KP1} and \cite{MP1} (for Weyl groups and orbit functions) as
well as in first papers of the authors \cite{HP2,PK}. Facts we
outline here are mostly to establish the notations and also for
the sake of completeness of this paper.

Let $G$ be a compact simply connected simple Lie group of rank
$n$. The Weyl group $W$ corresponding to $G$ is
generated by $n$ reflections and is
specified by its Coxeter-Dynkin diagram. We choose a
non-orthogonal basis $\Pi=\{\alpha_1,\dots,\alpha_n\}\subset
\R^n$  of the simple roots of $G$. Then $W\Pi$ is the
corresponding root system while
\begin{equation*}\label{q_lattice}
Q=\left\{\sum_{i=1}^n a_i\alpha_i\ |\,
a_i\in\Z\right\}=\Z\alpha_1+\dots+\Z \alpha_n
\end{equation*}
is the root lattice. Let $\langle\,,\,\rangle$ be a scalar
product on $\R^n$. Put
${\alpha^\vee}=2\alpha/\langle\alpha,\alpha\rangle\,$ for any
$\,\alpha\in\Pi$, name
${\Pi^\vee}=\{\alpha^\vee_1,\dots,\alpha^\vee_n\}\,$ the
coroot basis and all linear combinations $\,\sum_{i=1}^n
a_i\alpha^\vee_i,\ a_i\in\Z$ the coroot lattice~$Q^\vee$.
Relative length and angles between simple roots in $\Pi$ are
given by the elements of the Cartan matrix
$C=(c_{ij})^n_{i,j=1}$, namely
$c_{ij}=\langle\alpha_i,\alpha^\vee_j\rangle\,$ for
$i,j=1,\dots,n$. In addition to $\Pi$ and $\Pi^\vee$ we
also introduce the two other bases. The basis of fundamental
weights $\omega_1,\dots,\omega_n$ which is a dual basis of
$\Pi^\vee$, i.e. $$\langle
\omega_i,\alpha_j^\vee\rangle=\delta_{ij},\,\qquad
i,j=1,\dots,n,$$ while the basis of fundamental coweights
$\,\omega^\vee_1,\dots,\omega^\vee_n\,$ is dual to $\Pi$.
The $\omega-$basis and $\alpha-$basis are also related by
means of the Cartan matrix. In the matrix form
\begin{equation*}
\alpha =C\omega,\quad \quad \alpha^\vee=C^{\,T}\omega^\vee.
\end{equation*}
Rank $n$ lattices
$P=\Z\omega_1+\dots+\Z\omega_n$ and
$P^\vee=\Z\omega^\vee_1+\dots+\Z\omega^\vee_n$
are called correspondingly the weight lattice and coweight
lattice. The weight lattice $P$ is dual to the coroot lattice
$Q^\vee$, while $P^\vee$ is dual to $Q$.

We also define the set of dominant weights $P^+$ and the set
of strictly dominant weights $P^{++}$ as
\begin{equation*}\label{P^+}
P^+=\Z^{\geq 0}\omega_1+\Z^{\geq 0}\omega_2+\dots+\Z^{\geq
0}\omega_n \  \supset \  P^{++}=\Z^{> 0}\omega_1+\Z^{>
0}\omega_2+\dots+\Z^{> 0}\omega_n.
\end{equation*}

To any simple root $\alpha\in\Pi$ we associate the refection
$\,r_{\alpha}\,$ acting on $\,x\in\R^n\,$ as
$\,r_{\alpha}x=x-\langle \alpha,x\rangle\alpha^\vee$. Then the
finite Weyl group is generated by $r_i\equiv r_{\al_i}$, $i=1,\dots,n$.

Combining $W$ with the
translation group defined by $Q^\vee$ we obtain the
infinite affine Weyl group $W^{\mathrm{aff}}=Q^\vee\rtimes W$.
A fundamental region $F\subset\R^n$ for $W^{\mathrm{aff}}$ of
the simple compact Lie group $G$ of rank $n$ is the
simplex which is specified by $n+1$ vertices
$\{0,\tfrac{\omega^\vee_1}{m_1},\dots,\tfrac{\omega^\vee_n}{m_n}\}$,
where $m_1,\dots,m_n\,$ are the coefficients of the highest root
of $G$ relative to $\Pi$ -- see in \cite{HP1}.

Combining $W$ with the
translation group defined by $Q$ we obtain the
infinite dual affine Weyl group $\widehat{W}^{\mathrm{aff}}=Q\rtimes W$.
A fundamental region $F^\vee\subset\R^n$ for $\widehat{W}^{\mathrm{aff}}$
is the
simplex which is specified by $n+1$ vertices
$\{0,\tfrac{\omega_1}{m^\vee_1},\dots,\tfrac{\omega_n}{m^\vee_n}\}$,
where $m^\vee_1,\dots,m^\vee_n\,$ are the coefficients of the highest dual root of $G$ -- see in~\cite{HP1}.

For any $\lambda\in \R^n$
denote by $W(\lambda)$ its orbit with respect to the action of
$W$.

The even Weyl subgroup of $W$ is the set of elements of even
length, i.e.
\begin{equation}\label{def_even_group}
W^e=\langle
r_{i_1}\dots r_{i_k}|\,k {\mathrm{\ is\ even},\
}\,i_j\in\{1,\dots,n\}\rangle=\{w\in W\mid \det w = 1\}.
\end{equation}
It is a normal subgroup of $W$ of index $2$. For any
$\lambda\in \R^n$ its $W^e-$orbit is denoted by
$W^e(\lambda)$.
The even affine group is given by
\begin{equation*}
W^{\mathrm{aff}}_e= Q^\vee \rtimes W^e.
\end{equation*}
In the case of $W^{\mathrm{aff}}_e$ its fundamental region $F^e $ is the
union of the fundamental region $F$ of the original $W^{\mathrm{aff}}$ and arbitrary reflection $r_j\equiv r$ of its interior $r \intt(F)$:
\begin{equation}\label{funddome}
F^e=F\cup r�\intt(F).
\end{equation}
Similarly, we define a dual fundamental domain $F^{e\vee}$ of $\widehat{W}^{\mathrm{aff}}$ as
\begin{equation*}
F^{e\vee}=F^\vee\cup r�\intt(F^\vee).
\end{equation*}

\subsection{$E-$functions}\

The $E-$functions are orbit functions of the
symmetry group $W^e$:
\begin{equation*}
E_{\lambda}(x)=\sum_{\mu\in W^e(\lambda)}e^{2\pi
i\langle\,\mu,x\,\rangle},
\end{equation*}
for $\,x\in \R^n,\
\lambda\in P$. The $E-$functions are invariant under the action
both their corresponding even Weyl and even affine Weyl groups.

For any $\la\in \R^n$ we denote the order of the stabilizer
\begin{equation}\label{dla}
d^{e}_{\la}\equiv |\mathrm{Stab}_{W^e} (\la)|,\q \mathrm{Stab}_{W^e} (\la)=\set{w\in W^e}{w\la=\la}
\end{equation}
and introduce a different normalization of $E-$functions, namely
\begin{equation}\label{E}
 \Xi_\la (x)=d^{e}_{\la} E_{\la}(x)= \sum_{w\in W^e} e^{2 \pi \ii \sca{ w\la}{x}}.
\end{equation}

\subsection{Continuous orthogonality and continuous $E-$transforms}\

For any two weights from the set $P_e=P^+\cup r P^{++}$
corresponding $E-$functions are orthogonal on $F^e$
\begin{equation}\label{e_orthog}
\int_{{F}^e}\Xi_{\lambda}(x)\overline{\Xi_{\lambda'}(x)}\,dx=
|F^e|\,|W^e|\,d^{e}_{\la}\,\delta_{\lambda\lambda'}. \end{equation}
Here the overline denotes complex conjugation, $|F^e|$ the
volume of the domain $F^e$.  For the proof see~\cite{MP1}.

The $E-$functions determine a symmetrized Fourier series
expansions,
\begin{equation*}
f(x)=\sum_{\la\in P_{e}}c_{\la}\Xi_{\la}(x),\quad {\mathrm{
where}}\
c_{\la}=\frac{1}{|F^{e}||W^{e}|d^{e}_{\la}}\int_{F^{e}}f(x)\overline{\Xi_{\la}(x)}\,dx.
\end{equation*}

\subsection{Discrete orthogonality and discrete $E-$transforms}\

For an arbitrary natural number $M$, the discrete calculus of $E-$functions is performed over the intersection $F_M^e$ of the finite group $\frac{1}{M}P^{\vee}/Q^{\vee}$ with the fundamental domain $F^e$, $F^{e}_M\equiv\frac{1}{M}P^{\vee}/Q^{\vee}\cap F^{e}$. The finite set of dominant weights $\Lambda_M^e$ labeling orthogonal $E-$functions can be chosen as the intersection of the quotient group $P/MQ$ with the augmented dual fundamental domain $MF^{e\vee }$, $\Lambda^e_M\equiv M F^{e\vee} \cap P/MQ$.
 For $x\in \frac{1}{M}P^{\vee}/Q^{\vee}$ we denote the orbit and its order by
 \begin{equation}\label{epx}
W^e x=\set{wx\in \R^n/Q^{\vee} }{w\in W^e},\q \ep^e(x)\equiv |W^ex|.
\end{equation}
For $\la \in P/MQ$ we denote the order of the stabilizer
\begin{equation}\label{hla}
h^{e\vee}_{\la}\equiv |\mathrm{Stab}^{\vee}_e (\la)|,\q \mathrm{Stab}^{\vee}_e (\la)=\set{w\in W^e}{w\la=\la}.
\end{equation}
  For $\la,\la' \in\Lambda^e_M$ the discrete orthogonality relations hold \cite{HP2}
\begin{equation}\label{orthoE}
 \sum_{x\in F^e_M}\ep^e(x) \Xi_\la(x)\overline{\Xi_{\la'}(x)}=\det C \,\abs{W^e}M^n h^{e\vee}_\la \delta_{\la\la'}
\end{equation}
 and the discrete symmetrized $E-$functions expansion is given by                    \begin{equation}\label{disc_transform}
f(x)=\sum_{\la\in \Lambda_M^e}c_{\la}\Xi_{\la}(x),\quad {\mathrm{
where}}\
c_{\la}=\frac{1}{\det C \,\abs{W^e}M^n h^{e\vee}_\la}\sum_{x\in F^e_M}\ep^e(x) f(x)\overline{\Xi_{\la}(x)}.
\end{equation}
\begin{example}
In the case of the rank one group $A_1$
the Weyl group $W(A_1)$ has two elements $\{id,\,r\},$
where $r$ is the reflection in the origin. Then
$W^e(\Aone)=\{id\}$. The Cartan matrix of $\Aone$ is
$(2)$. Thus, we have $\al^\vee=\al$, $\om=\om^\vee$ and $\al=2\om$ with $\sca{\om}{\om}=1/2$. The weight lattice $P=\Z\om$, the dominant weights are
$P^+=\Z^{\geq 0}\om$ and $P_e=\Z\om$. For any $a\om \in \R$ it holds $W(a)=\{a,-a\}$, while
$\,W^e(a)=\{a\}$. Since $F(\Aone)= \set{x \om }{0\leq x\leq 1}$, we have from \eqref{funddome}
\begin{equation*}
F^e(\Aone)= \set{x \om }{-1< x\leq 1}.
\end{equation*}
 Therefore the corresponding $E-$function \eqref{E} for $a\om\in P_e$, $a\in \Z$ and $ x\om\in F^e$ is
\begin{equation*}
\Xi_a (x) =e^{2i\pi \sca{a\om}{x\om}}=e^{i\pi ax }.
\end{equation*}
One can easily check the property \eqref{e_orthog}, in particular for any
$\lambda=a\om,\lambda'=a'\om\in\Z\om$
\begin{equation*}\label{e_orthog_dim1}
\int_{F^e(\Aone)}\Xi_{\lambda}(z)\overline{\Xi_{\lambda'}(z)}\,dz=\tfrac{1}{\sqrt{2}}\int^1_{-1}e^{i\pi (a-a')x }\,dx=\sqrt{2}\delta_{\lambda \lambda'}.
\end{equation*}
 The discrete sets $F^e_M$ and $\Lambda^e_M$ have the form
 \begin{align*}
  F^e_M(A_1) =& \setb{\frac{s_1}{M}\om}{M<s_1\leq M} \\
  \Lambda^e_M(A_1) =& \set{t_1\om}{M<t_1\leq M}.
 \end{align*}
 The discrete orthogonality (\ref{orthoE}) has for $\la,\la'\in\Lambda^e_M(A_1) $ the form
 \begin{equation*}
 \sum_{z\in F^e_M(A_1)}\Xi_\la(z)\overline{\Xi_{\la'}(z)}=\sum_{s_1=-M+1}^{M} e^{\frac{i\pi (t_1-t_1')s_1}{M} } =2M \delta_{\la\la'}.
\end{equation*}
 \end{example}

\section{Even Weyl group of semisimple Lie groups}\label{EvenWeyl}
In this section we will define the even Weyl group for the
semisimple Lie groups. Let $G_1$, $G_2$ be two compact simply connected
simple Lie groups of rank $n_1$ and $n_2$
correspondingly. For $G_i\,,$ $\,i=1,2\,$ we denote by $W_i$ corresponding Weyl group,
by $C_i$ its Cartan matrix, by $P_i$ the weight lattice,
by $Q^\vee_i\,$ its coroot lattice
and by $F_i$ the corresponding fundamental region.
Then for the semisimple Lie group $G=G_1\times G_2$ of rank
$n_1+ n_2$ we obtain that the corresponding Weyl group is
$W=W_1\times W_2$, its Cartan matrix is
$C=\left(\begin{smallmatrix}
C_1&0\\0&C_2\end{smallmatrix}\right)$, the weight lattice
$P=P_1\times P_2\,$ both as lattices and as groups, analogously
the coroot lattice $Q^\vee=Q^\vee_1\times Q^\vee_2$, its
affine Weyl group $W^{\mathrm{aff}}=W_1^{\mathrm{aff}}\times W_2^{\mathrm
{aff}}=W\times Q^\vee$ and finally the fundamental region
$F$ is the cartesian product of $F_1$ and $F_2$.

There are two natural ways of defining even Weyl subgroup of
$W=W_1\times W_2$ (even more in the case when $G$ has more
then $2$ factors). For the first one continues the sequence of the
notions for semisimple group in the last paragraph: the even Weyl
subgroup of $W$ is a direct product of the corresponding even
Weyl groups $W^e_1$ and $W^e_2$, we will denote it by
\begin{equation}\label{e_group_type_1}
W^{ee}=W^e_1\times W^e_2.
\end{equation}
Its affine even group is $W^{\mathrm{aff}}_{ee}=(W_1)^{\mathrm
{aff}}_e\times(W_2)_e^{\mathrm{aff}}=W_1^e\times W_2^e\times
Q^\vee=W^{ee}\times Q^\vee$ and its fundamental region
$F^{ee}=F_1^e\times F_2^e$. The second possibility arises from
\eqref{def_even_group}, namely, we define the even subgroup
$W^e$ as the set of the elements of even length of $W$. In
this case its affine group $W_e^{\mathrm{aff}}=W^e\times Q^\vee$
and its fundamental region $F^e=F_1\times F_2 \cup \intt (r_1 F_1\times F_2)$. In spite of the fact that
the fundamental regions, where the expansion takes place as well
as expansion functions are different, for both cases we will have
both continuous and discrete orthogonality.

\subsection{Cartesian product $W^e(G_1)\times W^e(G_2)$}\

We consider the Cartesian product of two even Weyl groups $W^e(G_1)\equiv W^e_1
$ and $W^e(G_2)\equiv W^e_2$, i.e. $$W^{ee}=W^e_1\times W^e_2.$$ Since $|W_i^e|=\tfrac12|W_i|$ we obtain that
$|W^{ee}|=\tfrac14|W_1||W_2|$. Let $\lambda\in
\R^{n_1}$ and $\mu\in \R^{n_2}$ then for
$\nu=(\lambda,\mu)\in \R^{n_1+n_2}$ we obtain for $W^{ee}-$orbits and stabilizers
\begin{align*}
W^{ee}(\nu) &= W_1^e(\lambda)\times W_2^e(\mu) \\
  \mathrm{Stab}_{W^{ee}} (\nu)&= \mathrm{Stab}_{W_1^{e}}(\lambda)\times \mathrm{Stab}_{W_2^{e}}(\mu)
\end{align*}
and consequently
\begin{equation*}
 d^{ee}_{\nu}= |\mathrm{Stab}_{W^{ee}} (\nu)|=d^{e}_{\la}d^{e}_{\mu}.
\end{equation*}
Denote by $P_{ee}=(P_1)_e\times (P_2)_e$ and the normalized $E-$function corresponding to $W^{ee}$ by $\Xi^{ee}$. Then for any
$\nu=(\lambda,\mu)\in P_{ee}$ and
$x=(x_1,x_2)\in\R^{n_1+n_2}$ the corresponding $\Xi^{ee}-$function is of the form
\begin{equation}\label{e_function_first_type}
\Xi^{ee}_{\nu}(x)=\sum_{w\in W^{ee}}e^{2\pi
i\langle\,w \nu,x\,\rangle}=\sum_{w_1\in W^e_1} e^{2\pi
i\langle\,w_1 \la,x_1\,\rangle}\sum_{w_2\in W^e_2}e^{2\pi
i\langle\,w_2\mu,x_2\,\rangle}=\Xi_{\lambda}(x_1)\Xi_{\mu}(x_2).
\end{equation}
The functions $\Xi_{\lambda}$ and $\Xi_{\mu}$ are orbit
functions defined by $W^e_1$ and $W^e_2$.

\subsubsection{Continuous orthogonality and $E^{ee}-$transforms}\

Combining \eqref{e_function_first_type} with orthogonality
\eqref{e_orthog} for $E-$functions for $W_1^e$ and
$W_2^e$ we obtain the orthogonality for $\Xi^{ee}-$functions
defined by $W^{ee}$. For any
$\nu=(\lambda,\mu),\,\nu'=(\lambda',\mu')\in P_{ee}$
\begin{align}\label{orthog_product1}
\int_{F^{ee}}\Xi^{ee}_{\nu}(x)\overline{\Xi^{ee}_{\nu'}(x)}\,dx
&=\int_{F_1^e}\Xi_{\lambda}(x_1)\overline{\Xi_{\lambda'}(x_1)}\,dx_1
\int_{F_2^e}\Xi_{\mu}(x_2)\overline{\Xi_{\mu'}(x_2)}\,dx_2\nonumber \\
&=|F_1^e||W_1^e|d^{e}_{\la}\delta_{\lambda\lambda'}|F_2^e||W_2^e|d^{e}_{\mu}\delta_{\mu\mu'}=
|F^{ee}||W^{ee}|d^{ee}_{\nu}\delta_{\nu\nu'}.
\end{align}

Note that one can easily generalize this definition for the case
$G=G_1\times \dots\times G_k$. The even group is given by
$W^{ee}=W_1^e\times\dots\times W_k^e$. For any
$\nu=(\nu_1,\dots,\nu_k)\in P_{ee}\equiv(P_1)_e\times \dots\times (P_k)_e$
and $x=(x_1,\dots,x_k)\in\R^{n_1+\dots+n_k}$
\begin{equation}\label{multiple_e_type_1} \Xi^{ee}_{\nu}(x)=\Xi_{\nu_1}(x_1)\dots \Xi_{\nu_k}(x_k).
\end{equation}
The fundamental region is $F^{ee}=F_1^e\times\dots\times
F_k^e$ and the orthogonal relations for any
$\nu,\,\nu'\in P_{ee}$ are
\begin{equation}\label{orth_multiple_e_type_1}
\int_{F^{ee}}\Xi^{ee}_{\nu}(x)\overline{\Xi^{ee}_{\nu'}(x)}\, dx =
|F^{ee}||W^{ee}|d^{ee}_{\nu}\delta_{\nu\nu'},
\end{equation}
where
$d^{ee}_{\nu}=d^{e}_{\nu_1}\dots d^{e}_{\nu_k} .$

Let $f$ be a function defined on $F^{ee}$ then we may
expand $f$ as a sum of $\Xi^{ee}-$functions
\begin{equation}\label{cont_transform_type_1} f(x)=\sum_{\nu\in P_{ee}}c_{\nu}\Xi^{ee}_{\nu}(x),\quad {\mathrm{
where}}\
c_{\nu}=\frac{1}{|F^{ee}||W^{ee}|d^{ee}_{\nu}}\int_{F^{ee}}f(x)\overline{\Xi^{ee}_{\nu}(x)}\,dx.
\end{equation}

\subsubsection{Discrete orthogonality and $E^{ee}-$transforms}\

For arbitrary natural numbers $M_1,M_2$, the discrete calculus of $E^{ee}-$functions is performed over the intersection $F_{M_1M_2}^{ee}$ of the finite group $\frac{1}{M_1}P_1^{\vee}/Q_1^{\vee}\times\frac{1}{M_2}P_2^{\vee}/Q_2^{\vee}$ with the fundamental domain $F^{ee}=F_1^e\times F_2^e$, $$F^{ee}_{M_1M_2}\equiv\left(\frac{1}{M_1}P_1^{\vee}/Q_1^{\vee}\times\frac{1}{M_2}P_2^{\vee}/Q_2^{\vee}\right)\cap F^{ee}.$$ The finite set of dominant weights $\Lambda_M^{ee}$ labeling orthogonal $E^{ee}-$functions can be chosen as the intersection of the quotient group $P_1/M_1Q_1\times P_2/M_2Q_2$ with the augmented dual fundamental domain $M_1F^{\vee e}_1\times  M_2F^{\vee e}_2$, $$\Lambda^{ee}_{M_1M_2}=(P_1/M_1Q_1\times P_2/M_2Q_2) \cap (M_1F^{\vee e}_1\times  M_2F^{\vee e}_2).$$
 For $x=(x_1,x_2)\in \frac{1}{M_1}P_1^{\vee}/Q_1^{\vee}\times\frac{1}{M_2}P_2^{\vee}/Q_2^{\vee}$ we have for the orbit and its order
 \begin{equation*}
\ep^{ee}(x)\equiv |W^{ee}x|= \ep^{e}(x_1)\ep^{e}(x_2) .
\end{equation*}
For $\la=(\mu,\nu) \in P_1/M_1Q_1\times P_2/M_2Q_2$ we have for the order of the stabilizer
\begin{equation}\label{hlaee}
h^{ee\vee}_{\la }\equiv |\mathrm{Stab}^{\vee}_{ee} (\la)|=h^{e\vee}_{\mu }h^{e\vee}_{\nu }. \end{equation}
  For $\la,\la' \in\Lambda^{ee}_{M_1M_2}$ the discrete orthogonality relations hold
\begin{align}
 \sum_{x\in F^{ee}_{M_1M_2}}\ep^{ee}(x) \Xi^{ee}_\la(x)\overline{\Xi^{ee}_{\la'}(x)}=\sum_{x_1\in {F_1^{e}}_{M_1}}\ep^{e}(x_1) \Xi_\mu(x_1)\overline{\Xi_{\mu'}(x_1)}\sum_{x_2\in {F_2^{e}}_{M_2}}  \ep^{e}(x_2) \Xi_\nu(x_2)\overline{\Xi_{\nu'}(x_2)} \nonumber \\
 =\det C_1C_2 \,\abs{W_1^e}\abs{W_2^e}M_1^{n_1}M_2^{n_2} h^{e\vee}_\mu h^{e\vee}_\nu \delta_{\mu,\mu'} \delta_{\nu,\nu'}=\det C \,\abs{W^{ee}}M_1^{n_1}M_2^{n_2}  h^{ee\vee}_\la \delta_{\la\la'} \label{diskEE}
\end{align}
 and the discrete symmetrized $E-$functions expansion is given by                                                              \begin{equation}\label{disc_transform_ee}
f(x)=\sum_{\la\in \Lambda^{ee}_{M_1M_2}}\tilde c_{\la}\Xi^{ee}_{\nu}(x),\quad {\mathrm{
where}}\
\tilde c_{\la}=\frac{1}{\det C \,\abs{W^{ee}}M_1^{n_1}M_2^{n_2} h^{ee\vee}_\nu}\sum_{x\in F^{ee}_{M_1M_2}}\ep^{ee}(x) f(x)\overline{\Xi^{ee}_{\la}(x)}.
\end{equation}

\subsection{The full even group $W^e(G_1\times G_2)$}\

Let $\Pi_i$ be a basis of simple roots of $G_i$. Denote by
$R=\{r_{\alpha}\,|\,\alpha\in\Pi_1\cup\Pi_2\}.$ Then define the
even group $W^e$ as a set of the elements of even length of
$W=W_1\times W_2$
\begin{equation}\label{e_group_second_type}
W^e(G_1\times G_2)=\langle r_{i_1}\dots r_{i_k}\,|\, k{\ \mathrm{ is\
even\ } } r_{i_j}\in R \rangle = \{w\in W_1\times W_2\mid \det w = 1\}.
\end{equation} In this case $W^e(G)$ is a normal subgroup of
$W$ of index $2$, i.e.
$|W^e|=\tfrac12|W|=\tfrac12|W_1||W_2|$. Take any generating
reflection $\,r_i\,$ of $\,W_i,\,$ $\,i=1,2$, then
\begin{equation*}
W^e(G_1\times G_2)=W_1^e\times W_2^e \,\cup\, (r_1W_1^e\times r_2W_2^e)
\end{equation*} and for any
$\la=(\mu,\nu)\in \R^{n_1+n_2}$
\begin{equation*}
W^e(G_1\times G_2)(\la)
=\{\,(\eta,\theta),\,(r_1\eta,r_2\theta)\,|\,\eta\in
W_1^e(\mu),\, \theta\in W_2^e(\nu)\}.
\end{equation*}

The corresponding affine even group is $W_e^{\mathrm{aff}}(G_1\times
G_2)=W^e\rtimes Q^\vee$ and its fundamental domain is
\begin{equation}\label{funddom_e}
F^e=F_1\times F_2\cup \intt (r_1 F_1\times F_2)\subset
\R^{n_1+n_2}.
\end{equation} The
choice of $r_1$ in \eqref{funddom_e} is
arbitrary, we can choose any two adjacent copies of $F$. Finally
put $$P_{e}=(P^+_1\times P^+_2) \cup r_1 P^{++}_1\times P^{++}_2.$$

For any $\la=(\mu,\nu)\in P_e$ and
$x=(x_1,x_2)\in\R^{n_1+n_2}$ the corresponding normalized $E^e-$function

\begin{equation}\label{e_function_second_type}
\Xi^e_\la (x)= \sum_{w\in W^e} e^{2 \pi \ii \sca{ w\la}{x}}=\Xi_{\nu}(x_1)\Xi_{\mu}(x_2)+\Xi_{r_1\nu}(x_1)\Xi_{r_2\mu}(x_2).
\end{equation}
Note, that for any $\la\in P_e$ function $\Xi_{\la}$ is
$W_e^{\mathrm{aff}}$ invariant.
\subsubsection{Continuous orthogonality and $E^{e}-$transforms}\

The
orthogonality of $E^e-$functions
for any two weights from the set $P_e$ is given by

\begin{equation}\label{e_orthog_e}
\int_{{F}^e}\Xi^e_{\lambda}(x)\overline{\Xi^e_{\lambda'}(x)}\,dx=
|F^e|\,|W^e|\,d^{e}_{\la}\,\delta_{\lambda\lambda'}, \end{equation}
where $F^e$ is defined by (\ref{funddom_e}). The size of the stabilizer  $d^{e}_{\la}$ can be defined by (\ref{dla}), with $W^e$ of the form (\ref{e_group_second_type}). For the proof see \cite{MP1}.

The $E^e-$functions determine a symmetrized Fourier series
expansions,
\begin{equation}\label{cont_transform}
f(x)=\sum_{\la\in P_{e}}c_{\la}\Xi^e_{\la}(x),\quad {\mathrm{
where}}\
c_{\la}=\frac{1}{|F^{e}||W^{e}|d^{e}_{\la}}\int_{F^{e}}f(x)\overline{\Xi^e_{\la}(x)}\,dx.
\end{equation}

\subsubsection{Discrete orthogonality and discrete $E^e-$transforms}\

Taking an arbitrary natural number $M$, the discrete calculus of $E^e-$functions is performed over the intersection $F_M^e$ of the finite group $\frac{1}{M}P^{\vee}/Q^{\vee}$ with the fundamental domain $F^e$ of the form (\ref{funddom_e}), $$F^{e}_M\equiv\frac{1}{M}(P_1^{\vee}\times P^{\vee}_2)/(Q_1^{\vee}\times Q_2^{\vee})\cap F^{e}.$$ The finite set of dominant weights $\Lambda_M^e$ labeling orthogonal $E-$functions can be chosen using the dual fundamental domain
\begin{equation}\label{fundamental_domain_w_e}
F^{e\vee }=F^{\vee}_1\times F^{\vee}_2\cup \intt (r_1 F^{\vee}_1\times F^{\vee}_2). \end{equation}
 Then the intersection of the quotient group $P/MQ$ with the augmented domain $MF^{e\vee }$ forms the set $$\Lambda^e_M\equiv M F^{e\vee} \cap (P_1\times P_2)/[M(Q_1\times Q_2)].$$
 For $x\in \frac{1}{M}P^{\vee}/Q^{\vee}$ we define the orbit  and its size $\ep^e(x)$ by the equation (\ref{epx}) and similarly the stabilizer of $\la \in P/MQ$ and its size $h_\la^{e\vee}$ by (\ref{hla}).

Then for $\la,\la' \in\Lambda^e_M$ the discrete orthogonality relations hold
\begin{equation}\label{orthoEe}
 \sum_{x\in F^e_M}\ep^e(x) \Xi^e_\la(x)\overline{\Xi^e_{\la'}(x)}=\det C \,\abs{W^e}M^{n_1+n_2} h^{e\vee}_\la \delta_{\la\la'}
\end{equation}
 and the discrete symmetrized $E^e-$functions expansion is given by                                                                \begin{equation}\label{disc_transform_e}
f(x)=\sum_{\la\in \Lambda_M^e}c_{\la}\Xi^e_{\la}(x),\quad {\mathrm{
where}}\
c_{\la}=\frac{1}{\det C \,\abs{W^e}M^{n_1+n_2} h^{e\vee}_\la}\sum_{x\in F^e_M}\ep^e(x) f(x)\overline{\Xi^e_{\la}(x)}.
\end{equation}

\section{The $E-$transforms of semi-simple Lie groups of rank $\leq
3$}\label{Etrans}

In this section we deal with expansions of continuous and discrete functions on
both $F^{ee}$ and $F^e$ into series of continuous and discrete
$E-$functions and their inversion. There are four compact semi-simple Lie
groups of rank $3$, namely $\,\SU(2)\times \SU(2)\times \SU(2)$,
$\,\SU(2)\times\SU(3)\,$, $\,\SU(2)\times\O(5)\,$ and
$\,\SU(2)\times\G(2)$. We are using the following notation often
used to denote the corresponding Lie algebras:
$$
\Aone\leftrightarrow\SU(2)\,,\quad \Atwo\leftrightarrow
\SU(3)\,,\quad \Ctwo\leftrightarrow \O(5)\,,\quad
\G_2\leftrightarrow \G(2)\,.
$$

For a particular Lie group
$G$ in order to apply the transformations \eqref{cont_transform_type_1}, \eqref{disc_transform_ee} and the transformations \eqref{cont_transform}, \eqref{disc_transform_e} we need the
following information:
\begin{enumerate}[1.]
\item the fundamental regions $F^{ee}$ and $F^{e}$ and their volumes
\item the infinite sets of weights $P_{ee}$ and $P_e$
\item the coefficients $d_\la^{ee}$ and $d_\la^{e}$
\item the explicit form of $E^{ee}-$ and $E^{e}-$functions
\item the discrete sets $F_{M_1M_2}^{ee}$ and $F_M^{e}$
\item the finite sets of weights $\Lambda^{ee}_{M_1M_2}$ and $\Lambda^e_M$
\item the coefficients $\ep^e(x)$, $\ep^{ee}(x)$ and $h^{e\vee}_\la$, $h^{ ee\vee}_\la$, $\det C$.
\end{enumerate}
\subsection{The $E-$transforms of $\Aone\times\Aone$}\

Relative length and angles between the simple roots and
fundamental weights of $\Aone\times\Aone$ are given in terms
of Cartan matrix
$C=\left(\begin{smallmatrix}2&0\\0&2\end{smallmatrix}\right)$, $\det C= 4$.

\begin{enumerate}[1.]
\item $W^{ee}(\Aone\times\Aone)$. The $W^{ee}-$orbit for the generic point $a\om_1+b\om_2\in \R^2 $ is given by
$$ W^{ee}{(a,b)}=
\{(a,b)\}$$
and $|W^{ee}|=1$.
The fundamental region $F^{ee}$ is given by
\begin{align*}
F^{ee}(\Aone\times\Aone)=&\{x\omega^\vee_1+y\om^\vee_2\mid
-1< x,y\leq 1\}
\end{align*}
and its volume is $| F^{ee}|=2$.
We have the weight lattice $P_{ee}$ of the form
\begin{align*}
P_{ee}(\Aone\times\Aone)=&\{a\om_1+b\om_2  \mid a,b\in\Z\}
\end{align*}
Then for any
$a\omega_1+b\omega_2\in P_{ee}$ the corresponding
$E^{ee}-$function at point
$x\omega^\vee_1+y\omega^\vee_2$ is
\begin{align*}
\Xi^{ee}_{(a,b)}(x,y)  = e^{\ii\pi (ax+by)}.
\end{align*}

The contour plots of some lowest $\Xi^e$ functions are plotted in Figure \ref{Aoneepic}.
\begin{figure}
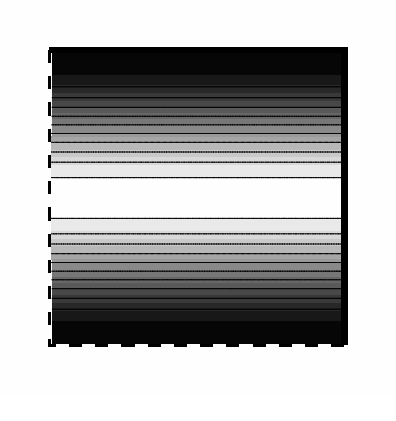\vspace*{0pt}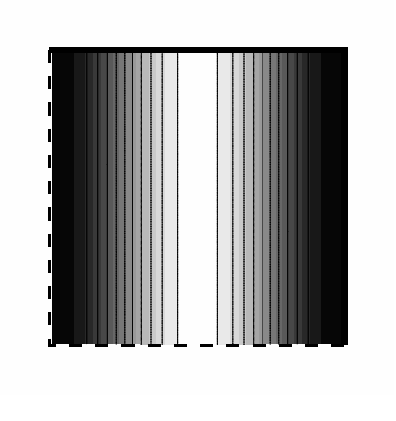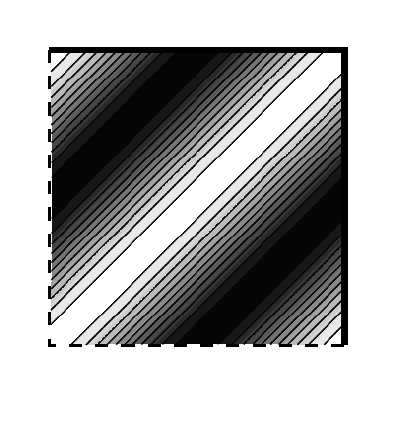\\
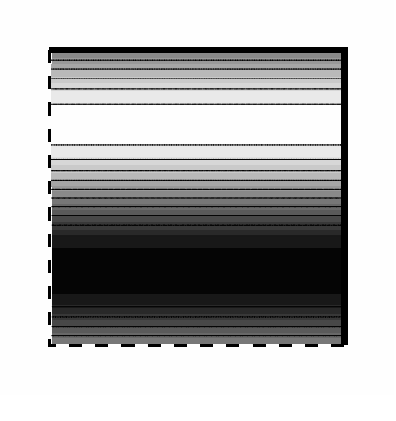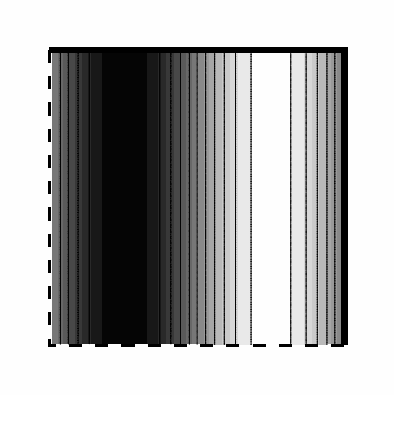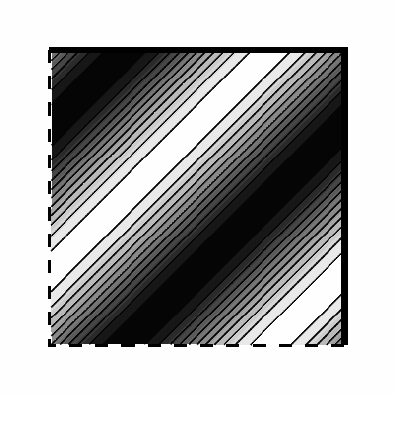\caption{The contour plots of $\Xi^{ee}-$functions of $\Aone\times\Aone$ plotted over the fundamental domain $F^{ee}$.}\label{Aoneepic}
 \end{figure}

The coefficients $d^{ee}_{(a,b)}$ in continuous orthogonality  relations (\ref{orthog_product1}) are all equal to $1$.

The discrete grid $F^{ee}_{M_1M_2}$ has the explicit form
\begin{align*}
F^{ee}_{M_1M_2}&=\setb{\frac{s_1}{M_1}\om^\vee_1+\frac{s_2}{M_2}\om^\vee_2}{s_i\in \Z, -M_i<s_i\leq M_i,\,i=1,2}
\end{align*}
and the corresponding grid $\Lambda^{ee}_{M_1M_2}$ of weights has the following form
\begin{align*}
\Lambda^{ee}_{M_1M_2}=& \set{t_1\om_1+t_2\om_2}{ t_i\in \Z,\,  -M_i<t_i\leq M_i,\, i=1,2}.
\end{align*}

The discrete orthogonality relations of the functions $\Xi^{ee}$ are of the form (\ref{diskEE}) with the resulting normalization coefficient equal to $4M_1M_2h^{ee\vee}_\la $. The coefficients $\ep^{ee}(x)$ and $h^{ee\vee}_\la$ in (\ref{diskEE}) are all equal to $1$.

\item $W^e(\Aone\times\Aone).$ The $W^{e}-$orbit for the generic point $a\om_1+b\om_2\in \R^2 $ is given by
$$ W^{e}{(a,b)}=\{(a, b),(-a,- b)\}$$
and $|W^{e}|=2$.
The fundamental region $F^{e}$ is given by
\begin{align*}
F^{e}(\Aone\times\Aone)=&\{x\omega^\vee_1+y\om^\vee_2\mid
0\leq x,y\leq 1 \}\cup\\ & \cup \{-x\omega^\vee_1+y\om^\vee_2\mid
0< x,y< 1\}
\end{align*}
and its volume is $| F^{e}|=1$.
We have the weight lattice $P_{e}$ of the form
\begin{align*}
P_{e}(\Aone\times\Aone)=&\{a\om_1+b\om_2  \mid a, b\in \Z^{\geq 0}\} \cup \\ &\cup \{-a\om_1+b\om_2\mid a, b\in \N\}.
\end{align*}
Then for any
$a\omega_1+b\omega_2\in P_{e}$ the corresponding
$E^{e}-$function at point
$x\omega^\vee_1+y\omega^\vee_2$ is
\begin{align*}
\Xi^e_{(a,b)}(x,y) & =2 \cos(\pi (a x+b y)).
\end{align*}
The contour plots of some lowest $\Xi^e$ functions are plotted in Figure \ref{Aonepic}.

\begin{figure}
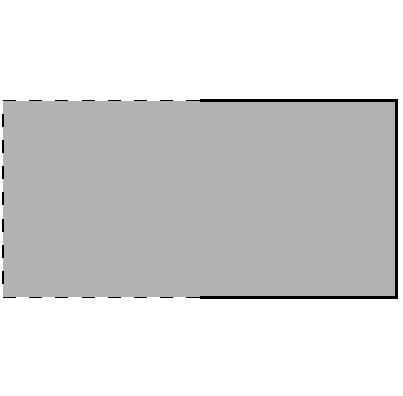\vspace*{-7pt}\hspace{20pt}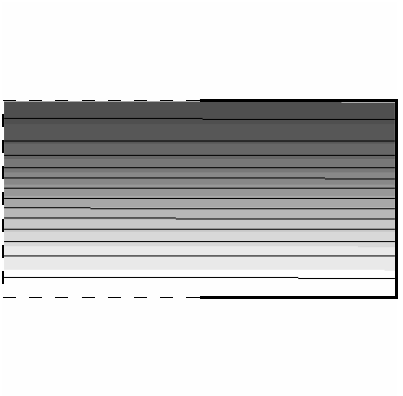\hspace{20pt}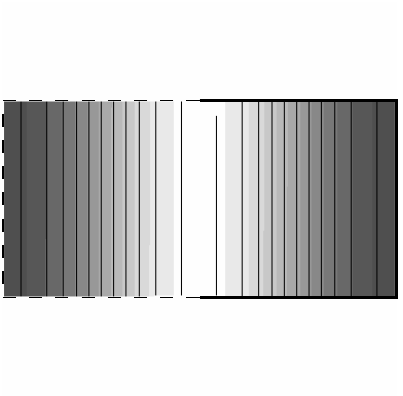\\
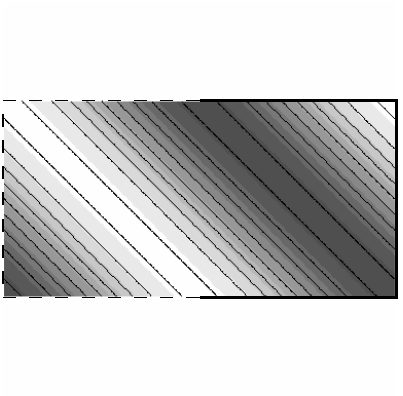\hspace{20pt}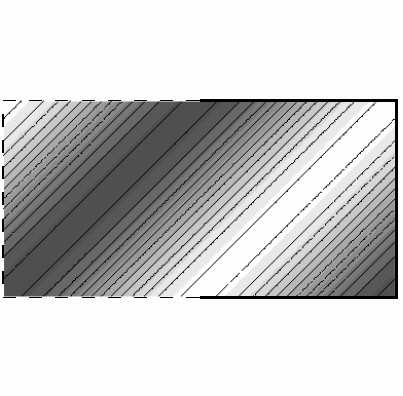\hspace{20pt}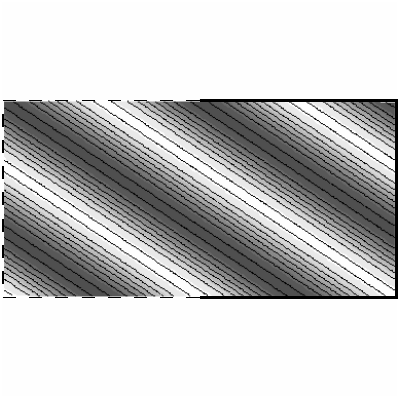
\caption{The contour plots of $\Xi^e-$functions of $\Aone\times\Aone$ plotted over the fundamental domain $F^e$.}\label{Aonepic}
\end{figure}

The coefficients $d^{e}_{(a,b)}$ in continuous orthogonality  relations (\ref{e_orthog_e}) have values in Table \ref{TabAone2}.

The discrete grid $F^{e}_{M}$ has the explicit form

\begin{align*}
F^{e}_{M}=&\setb{\frac{s_1}{M}\om^\vee_1+\frac{s_2}{M}\om^\vee_2}{s_1, s_2\in \Z^{\geq 0},  s_1,s_2\leq M   }\cup \\ &\cup \setb{-\frac{s_1}{M}\om^\vee_1+\frac{s_2}{M}\om^\vee_2}{s_1, s_2\in \N,  s_1,s_2< M}
\end{align*}
and the corresponding grid $\Lambda^{e}_{M}$ of weights has the following form
\begin{align*}
\Lambda^{e}_{M}=& \set{t_1\om_1+t_2\om_2}{ t_1, t_2\in \Z^{\geq 0}, t_1,t_2\leq M  } \cup \\ &\cup  \set{-t_1\om_1+t_2\om_2}{ t_1, t_2\in \N,\, t_1, t_2< M  }.
\end{align*}

The discrete orthogonality relations of the functions $\Xi^{e}$ are of the form (\ref{orthoEe}) with the resulting normalization coefficient equal to $8M^2h^{e\vee}_\la $. We label each point $x=\frac{s_1}{M}\om^\vee_1+\frac{s_2}{M}\om^\vee_2\in F^{e}_{M}$ by coordinates $[s_0,s_1,s_0',s_2]$ such that $s_0+s_1=M$, $s_0'+s_2=M$. Similarly, we label each point from $\la=t_1\om_1+t_2\om_2\in \Lambda^{e}_{M}$ by coordinates $[t_0,t_1,t_0',t_2]$ such that $t_0+t_1=M$, $t_0'+t_2=M$. The coefficients $\ep^{e}(x)$ and $h^{e\vee}_\la$ in (\ref{orthoEe}) are listed in Table \ref{TabAone2}.
\end{enumerate}

\subsection{The $E-$transforms of $\Aone\times\Atwo$}\

Relative length and angles between the simple roots and
fundamental weights of $\Aone\times\Atwo$ are given in terms
of Cartan matrix
$C=\left(\begin{smallmatrix}2&0&0\\0&2&-1\\0&-1&2\end{smallmatrix}\right) $, $\det C= 6$.

\begin{enumerate}[1.]
\item $W^{ee}(\Aone\times\Atwo)$. The $W^{ee}-$orbit for the generic point $a\om_1+b\om_2+c\om_3\in \R^3 $ is given by
$$ W^{ee}{(a,b,c)}=
\{(a,b,c),(a,c,-b-c),(a,-b-c,b)\}$$
and $|W^{ee}|=3$.
The fundamental region $F^{ee}$ is given by
\begin{align*}
F^{ee}(\Aone\times\Atwo)=&\{x\omega^\vee_1+y\om^\vee_2+z\om^\vee_3\mid
-1< x\leq 1,\, y,z\geq 0,\,  y+z\leq 1\}\cup\\ & \cup \{x\omega^\vee_1-y\om^\vee_2+(y+z)\om^\vee_3\mid
-1< x\leq 1,\, \, y,z> 0,\,  y+z< 1\}
\end{align*}
and its volume is $| F^{ee}|=2/\sqrt{6}$.
We have the weight lattice $P_{ee}$ of the form
\begin{align*}
P_{ee}(\Aone\times\Atwo)=&\{a\om_1+b\om_2+c\om_3  \mid a\in\Z ,\, b,c\in \Z^{\geq 0}\} \cup \\ &\cup \{a\om_1-b\om_2+(b+c)\om_3\mid a\in\Z ,\, b,c\in \N\}.
\end{align*}
Then for any
$a\omega_1+b\omega_2+c\omega_3\in P_{ee}$ the corresponding
$E^{ee}-$function at point
$x\omega^\vee_1+y\omega^\vee_2+z\omega^\vee_3$ is
\begin{align*}
\Xi^{ee}_{(a,b,c)}(x,y,z)  = e^{i\pi ax}(e^{\tfrac{2\pi
         i}3((2b+c)y+(b+2c)z)}+e^{-\tfrac{2\pi i}
         3((y+2z)b+(z-y)c)}+e^{-\tfrac{2\pi
         i}3((y-z)b+(2y+z)c)})
\end{align*}
and the coefficients $d^{ee}_{(a,b,c)}$ in continuous orthogonality  relations (\ref{orthog_product1}) have values in Table \ref{TabContee}.

The discrete grid $F^{ee}_{M_1M_2}$ has the explicit form
\begin{align*}
F^{ee}_{M_1M_2}=&\setb{\frac{s_1}{M_1}\om^\vee_1+\frac{s_2}{M_2}\om^\vee_2+\frac{s_3}{M_2}\om^\vee_3}{s_1\in \Z,\, s_2,s_3\in \Z^{\geq 0}, -M_1<s_1\leq M_1,\, s_2+s_3\leq M_2   }\cup \\ &\cup \setb{\frac{s_1}{M_1}\om^\vee_1-\frac{s_2}{M_2}\om^\vee_2+\frac{s_2+s_3}{M_2}\om^\vee_3}{s_1\in \Z,\, s_2,s_3\in \N, -M_1<s_1\leq M_1,\, s_2+s_3< M_2   }
\end{align*}
and the corresponding grid $\Lambda^{ee}_{M_1M_2}$ of weights has the following form
\begin{align*}
\Lambda^{ee}_{M_1M_2}=& \set{t_1\om_1+t_2\om_2+t_3\om_3}{ t_1\in \Z,\, t_2,t_3\in \Z^{\geq 0}, -M_1<t_1\leq M_1,\, t_2+t_3\leq M_2  } \cup \\ &\cup  \set{t_1\om_1-t_2\om_2+(t_2+t_3)\om_3}{ t_1\in \Z,\, t_2,t_3\in \N, -M_1<t_1\leq M_1,\, t_2+t_3< M_2  }.
\end{align*}

The discrete orthogonality relations of the functions $\Xi^{ee}$ are of the form (\ref{diskEE}). We label each point $x=\frac{s_1}{M_1}\om^\vee_1+\frac{s_2}{M_2}\om^\vee_2+\frac{s_3}{M_2}\om^\vee_3\in F^{ee}_{M_1M_2}$ by coordinates $[s_0,s_1,s_0',s_2,s_3]$ such that $s_0+s_1=M_1$ and $s_0'+s_2+s_3=M_2$. Similarly, we label each point from $\la=t_1\om_1+t_2\om_2+t_3\om_3\in \Lambda^{ee}_{M_1M_2}$ by coordinates $[t_0,t_1,t_0',t_2,t_3]$ such that $t_0+t_1=M_1$ and $t_0'+t_2+t_3=M_2$. The coefficients $\ep^{ee}(x)$ and $h^{ee\vee}_\la$ in (\ref{diskEE}) are listed in Table \ref{TabDiskee}.

\item $W^e(\Aone\times\Atwo).$ The $W^{e}-$orbit for the generic point $a\om_1+b\om_2+c\om_3\in \R^3 $ is given by
$$ W^{e}{(a,b,c)}=
\{(a,b,c),(a,c,-b-c),(a,-b-c,b),\\ (-a,-b,b+c),(-a,b+c,-c),(-a,-c,-b)\}$$
and $|W^{e}|=6$.
The fundamental region $F^{e}$ is given by
\begin{align*}
F^{e}(\Aone\times\Atwo)=&\{x\omega^\vee_1+y\om^\vee_2+z\om^\vee_3\mid
0\leq x\leq 1,\, y,z\geq 0,\,  y+z\leq 1\}\cup\\ & \cup \{x\omega^\vee_1-y\om^\vee_2+(y+z)\om^\vee_3\mid
0< x< 1,\, \, y,z> 0,\,  y+z< 1\}
\end{align*}
and its volume is $| F^{e}|=1/\sqrt{6}$.
We have the weight lattice $P_{e}$ of the form
\begin{align*}
P_{e}(\Aone\times\Atwo)=&\{a\om_1+b\om_2+c\om_3  \mid a,b,c\in\Z^{\geq 0} \} \cup \\ &\cup \{a\om_1-b\om_2+(b+c)\om_3\mid a,b,c\in\N \}.
\end{align*}
Then for any
$a\omega_1+b\omega_2+c\omega_3\in P_{e}$ the corresponding
$E^{e}-$function at point
$x\omega^\vee_1+y\omega^\vee_2+z\omega^\vee_3$ is
\begin{align*}
\Xi^{e}_{(a,b,c)}(x,y,z) & =e^{i\pi
ax}(e^{\tfrac{2\pi i}3((2b+c)y+(b+2c)z)}+e^{-\tfrac{2\pi i}3((b-c)y+(2b+c)z)}+e^{-\tfrac{2\pi i}3((b+2c)y+(c-b)z)})+\\
& e^{-i\pi ax}(e^{\tfrac{2\pi i}3((c-b)y+(b+2c)z)}+e^{\tfrac{2\pi
i}3((2b+c)y+(b-c)z)}+e^{-\tfrac{2\pi i}3((2c+b)y+(c+2b)z)})
\end{align*}
and the coefficients $d^{e}_{(a,b,c)}$ in continuous orthogonality  relations (\ref{e_orthog_e}) have values in Table \ref{TabConte}.

The discrete grid $F^{e}_{M}$ has the explicit form

\begin{align*}
F^{e}_{M}=&\setb{\frac{s_1}{M}\om^\vee_1+\frac{s_2}{M}\om^\vee_2+\frac{s_3}{M}\om^\vee_3}{s_1, s_2,s_3\in \Z^{\geq 0},  s_1\leq M,\, s_2+s_3\leq M   }\cup \\ &\cup \setb{\frac{s_1}{M}\om^\vee_1-\frac{s_2}{M}\om^\vee_2+\frac{s_2+s_3}{M}\om^\vee_3}{s_1, s_2,s_3\in \N,  s_1< M,\, s_2+s_3< M}
\end{align*}
and the corresponding grid $\Lambda^{e}_{M}$ of weights has the following form
\begin{align*}
\Lambda^{e}_{M}=& \set{t_1\om_1+t_2\om_2+t_3\om_3}{ t_1, t_2,t_3\in \Z^{\geq 0}, t_1\leq M,\, t_2+t_3\leq M  } \cup \\ &\cup  \set{t_1\om_1-t_2\om_2+(t_2+t_3)\om_3}{ t_1, t_2,t_3\in \N, t_1< M,\, t_2+t_3< M  }.
\end{align*}

The discrete orthogonality relations of the functions $\Xi^{e}$ are of the form (\ref{orthoEe}). We label each point $x=\frac{s_1}{M}\om^\vee_1+\frac{s_2}{M}\om^\vee_2+\frac{s_3}{M}\om^\vee_3\in F^{e}_{M}$ by coordinates $[s_0,s_1,s_0',s_2,s_3]$ such that $s_0+s_1=M$ and $s_0'+s_2+s_3=M$. Similarly, we label each point from $\la=t_1\om_1+t_2\om_2+t_3\om_3\in \Lambda^{e}_{M}$ by coordinates $[t_0,t_1,t_0',t_2,t_3]$ such that $t_0+t_1=M$ and $t_0'+t_2+t_3=M$. The coefficients $\ep^{e}(x)$ and $h^{e\vee}_\la$ in (\ref{orthoEe}) are listed in Table \ref{TabDiske}.
\end{enumerate}

\subsection{The $E-$transforms of $\Aone\times\Ctwo$}\

Relative length and angles between the simple roots and
fundamental weights of $\,\Aone\times\Ctwo\,$ are given in terms
of Cartan matrix
$\,C=\left(\begin{smallmatrix}2&0&0\\0&2&-1\\0&-2&2\end{smallmatrix}\right)$, $\det C= 4$.

\begin{enumerate}[1.]
\item $W^{ee}(\Aone\times\Ctwo)$. The $W^{ee}-$orbit for the generic point $a\om_1+b\om_2+c\om_3\in \R^3 $ is given by
$$ W^{ee}{(a,b,c)}=
\{(a,\pm b,\pm c),(a,\pm (2c+b),\mp (b+c))\}$$
and $|W^{ee}|=4$.
The fundamental region $F^{ee}$ is given by
\begin{align*}
F^{ee}(\Aone\times\Ctwo)=&\{x\omega^\vee_1+y\om^\vee_2+z\om^\vee_3\mid
-1< x\leq 1,\, y,z\geq 0,\,  2y+z\leq 1\}\cup\\ & \cup \{x\omega^\vee_1-y\om^\vee_2+(z+2y)\om^\vee_3\mid
-1< x\leq 1,\, \, y,z> 0,\,  2y+z< 1\}
\end{align*}
and its volume is $| F^{ee}|=1/\sqrt{2}$.
We have the weight lattice $P_{ee}$ of the form
\begin{align*}
P_{ee}(\Aone\times\Ctwo)=&\{a\om_1+b\om_2+c\om_3  \mid a\in\Z ,\, b,c\in \Z^{\geq 0}\} \cup \\ &\cup \{a\om_1-b\om_2+(b+c)\om_3\mid a\in\Z ,\, b,c\in \N\}.
\end{align*}
Then for any
$a\omega_1+b\omega_2+c\omega_3\in P_{ee}$ the corresponding
$E^{ee}-$function at point
$x\omega^\vee_1+y\omega^\vee_2+z\omega^\vee_3$ is
\begin{align*}
\Xi^{ee}_{(a,b,c)}(x,y,z)  = 2e^{\ii\pi
ax}(\cos(\pi((2b+2c)y+(b+2c)z))+\cos(\pi(2cy-bz)))
\end{align*}
and the coefficients $d^{ee}_{(a,b,c)}$ in continuous orthogonality  relations (\ref{orthog_product1}) have values in Table \ref{TabContee}.

The discrete grid $F^{ee}_{M_1M_2}$ has the explicit form
\begin{align*}
F^{ee}_{M_1M_2}=&\setb{\frac{s_1}{M_1}\om^\vee_1+\frac{s_2}{M_2}\om^\vee_2+\frac{s_3}{M_2}\om^\vee_3}{s_1\in \Z,\, s_2,s_3\in \Z^{\geq 0}, -M_1<s_1\leq M_1,\, 2s_2+s_3\leq M_2   }\cup \\ &\cup \setb{\frac{s_1}{M_1}\om^\vee_1-\frac{s_2}{M_2}\om^\vee_2+\frac{s_3+2s_2}{M_2}\om^\vee_3}{s_1\in \Z,\, s_2,s_3\in \N, -M_1<s_1\leq M_1,\, 2s_2+s_3< M_2   }
\end{align*}
and the corresponding grid $\Lambda^{ee}_{M_1M_2}$ of weights has the following form
\begin{align*}
\Lambda^{ee}_{M_1M_2}=& \set{t_1\om_1+t_2\om_2+t_3\om_3}{ t_1\in \Z,\, t_2,t_3\in \Z^{\geq 0}, -M_1<t_1\leq M_1,\, t_2+2t_3\leq M_2  } \cup \\ &\cup  \set{t_1\om_1-t_2\om_2+(t_2+t_3)\om_3}{ t_1\in \Z,\, t_2,t_3\in \N, -M_1<t_1\leq M_1,\, t_2+2t_3< M_2  }.
\end{align*}

The discrete orthogonality relations of the functions $\Xi^{ee}$ are of the form (\ref{diskEE}). We label each point $x=\frac{s_1}{M_1}\om^\vee_1+\frac{s_2}{M_2}\om^\vee_2+\frac{s_3}{M_2}\om^\vee_3\in F^{ee}_{M_1M_2}$ by coordinates $[s_0,s_1,s_0',s_2,s_3]$ such that $s_0+s_1=M_1$ and $s_0'+2s_2+s_3=M_2$. Similarly, we label each point from $\la=t_1\om_1+t_2\om_2+t_3\om_3\in \Lambda^{ee}_{M_1M_2}$ by coordinates $[t_0,t_1,t_0',t_2,t_3]$ such that $t_0+t_1=M_1$ and $t_0'+t_2+2t_3=M_2$. The coefficients $\ep^{ee}(x)$ and $h^{ee\vee}_\la$ in (\ref{diskEE}) are listed in Table \ref{TabDiskee}.

\item $W^e(\Aone\times\Ctwo).$ The $W^{e}-$orbit for the generic point $a\om_1+b\om_2+c\om_3\in \R^3 $ is given by
$$ W^{e}{(a,b,c)}=
\{(a,\pm b,\pm c),(a,\pm (2c+b),\mp (b+c)), (-a,\pm b,\mp
(b+c)), (-a,\pm(2c+b),\mp c)\}$$
and $|W^{e}|=8$.
The fundamental region $F^{e}$ is given by
\begin{align*}
F^{e}(\Aone\times\Ctwo)=&\{x\omega^\vee_1+y\om^\vee_2+z\om^\vee_3\mid
0\leq x\leq 1,\, y,z\geq 0,\,  2y+z\leq 1\}\cup\\ & \cup \{x\omega^\vee_1-y\om^\vee_2+(z+2y)\om^\vee_3\mid
0< x< 1,\, \, y,z> 0,\,  2y+z< 1\}
\end{align*}
and its volume is $| F^{e}|=\sqrt{2}/4$.
We have the weight lattice $P_{e}$ of the form
\begin{align*}
P_{e}(\Aone\times\Ctwo)=&\{a\om_1+b\om_2+c\om_3  \mid a,b,c\in\Z^{\geq 0} \} \cup \\ &\cup \{a\om_1-b\om_2+(b+c)\om_3\mid a,b,c\in\N \}.
\end{align*}
Then for any
$a\omega_1+b\omega_2+c\omega_3\in P_{e}$ the corresponding
$E^{e}-$function at point
$x\omega^\vee_1+y\omega^\vee_2+z\omega^\vee_3$ is
\begin{align*}
\Xi^{e}_{(a,b,c)}(x,y,z) & =2e^{i\pi
ax}(\cos(\pi((2b+2c)y+(b+2c)z))+\cos(\pi(2cy-bz)))+\\ & 2e^{-i\pi
ax}(\cos(\pi(2cy+(b+2c)z))+\cos(\pi(bz+(2b+2c)y))),
\end{align*}
and the coefficients $d^{e}_{(a,b,c)}$ in continuous orthogonality  relations (\ref{e_orthog_e}) have values in Table \ref{TabConte}.

The discrete grid $F^{e}_{M}$ has the explicit form

\begin{align*}
F^{e}_{M}=&\setb{\frac{s_1}{M}\om^\vee_1+\frac{s_2}{M}\om^\vee_2+\frac{s_3}{M}\om^\vee_3}{s_1, s_2,s_3\in \Z^{\geq 0},  s_1\leq M,\, 2s_2+s_3\leq M   }\cup \\ &\cup \setb{\frac{s_1}{M}\om^\vee_1-\frac{s_2}{M}\om^\vee_2+\frac{s_3+2s_2}{M}\om^\vee_3}{s_1, s_2,s_3\in \N,  s_1< M,\, 2s_2+s_3< M}
\end{align*}
and the corresponding grid $\Lambda^{e}_{M}$ of weights has the following form
\begin{align*}
\Lambda^{e}_{M}=& \set{t_1\om_1+t_2\om_2+t_3\om_3}{ t_1, t_2,t_3\in \Z^{\geq 0}, t_1\leq M,\, t_2+2t_3\leq M  } \cup \\ &\cup  \set{t_1\om_1-t_2\om_2+(t_2+t_3)\om_3}{ t_1, t_2,t_3\in \N, \,t_1< M,\, t_2+2t_3< M  }.
\end{align*}

The discrete orthogonality relations of the functions $\Xi^{e}$ are of the form (\ref{orthoEe}). We label each point $x=\frac{s_1}{M}\om^\vee_1+\frac{s_2}{M}\om^\vee_2+\frac{s_3}{M}\om^\vee_3\in F^{e}_{M}$ by coordinates $[s_0,s_1,s_0',s_2,s_3]$ such that $s_0+s_1=M$ and $s_0'+2s_2+s_3=M$. Similarly, we label each point from $\la=t_1\om_1+t_2\om_2+t_3\om_3\in \Lambda^{e}_{M}$ by coordinates $[t_0,t_1,t_0',t_2,t_3]$ such that $t_0+t_1=M$ and $t_0'+t_2+2t_3=M$. The coefficients $\ep^{e}(x)$ and $h^{e\vee}_\la$ in (\ref{orthoEe}) are listed in Table \ref{TabDiske}.
\end{enumerate}

\subsection{The $E-$transforms of $\Aone\times\Gtwo$}\

Relative length and angles between the simple roots and
fundamental weights of $\,\Aone\times\Ctwo\,$ are given in terms
of Cartan matrix
$C=\left(\begin{smallmatrix}2&0&0\\0&2&-3\\0&-1&2\end{smallmatrix}\right)$, $\det C= 2$.

\begin{enumerate}[1.]
\item $W^{ee}(\Aone\times\Gtwo)$. The $W^{ee}-$orbit for the generic point $a\om_1+b\om_2+c\om_3\in \R^3 $ is given by
$$ W^{ee}{(a,b,c)}=
\{(a,\pm b,\pm c),(a,\pm (2b+c),\mp (3b+c)),(a,\mp(b+c),\pm(3b+2c))\}$$
and $|W^{ee}|=6$.
The fundamental region $F^{ee}$ is given by
\begin{align*}
F^{ee}(\Aone\times\Gtwo)=&\{x\omega^\vee_1+y\om^\vee_2+z\om^\vee_3\mid
-1< x\leq 1,\, y,z\geq 0,\,  2y+3z\leq 1\}\cup\\ & \cup \{x\omega^\vee_1-y\om^\vee_2+(y+z)\om^\vee_3\mid
-1< x\leq 1,\, \, y,z> 0,\,  2y+3z< 1\}
\end{align*}
and its volume is $| F^{ee}|=\sqrt{6}/6$.
We have the weight lattice $P_{ee}$ of the form
\begin{align*}
P_{ee}(\Aone\times\Ctwo)=&\{a\om_1+b\om_2+c\om_3  \mid a\in\Z ,\, b,c\in \Z^{\geq 0}\} \cup \\ &\cup \{a\om_1-b\om_2+(c+3b)\om_3\mid a\in\Z ,\, b,c\in \N\}.
\end{align*}
Then for any
$a\omega_1+b\omega_2+c\omega_3\in P_{ee}$ the corresponding
$E^{ee}-$function at point
$x\omega^\vee_1+y\omega^\vee_2+z\omega^\vee_3$ is
\begin{align*}
\Xi^{ee}_{(a,b,c)}(x,y,z)  &= 2e^{i\pi ax}(\cos(2\pi((2b+c)y+(3b+2c)z))+2\cos(2\pi(bx+(3b+c)z))\\
&  \ +2\cos(2\pi((b+c)y+cz)))
\end{align*}
and the coefficients $d^{ee}_{(a,b,c)}$ in continuous orthogonality  relations (\ref{orthog_product1}) have values in Table \ref{TabContee}.

The discrete grid $F^{ee}_{M_1M_2}$ has the explicit form
\begin{align*}
F^{ee}_{M_1M_2}&=\setb{\frac{s_1}{M_1}\om^\vee_1+\frac{s_2}{M_2}\om^\vee_2+\frac{s_3}{M_2}\om^\vee_3}{s_1\in \Z,\, s_2,s_3\in \Z^{\geq 0}, -M_1<s_1\leq M_1,\, 2s_2+3s_3\leq M_2   }\cup \\ &\cup \setb{\frac{s_1}{M_1}\om^\vee_1-\frac{s_2}{M_2}\om^\vee_2+\frac{s_2+s_3}{M_2}\om^\vee_3}{s_1\in \Z,\, s_2,s_3\in \N, -M_1<s_1\leq M_1,\, 2s_2+3s_3< M_2   }
\end{align*}
and the corresponding grid $\Lambda^{ee}_{M_1M_2}$ of weights has the following form
\begin{align*}
\Lambda^{ee}_{M_1M_2}=& \set{t_1\om_1+t_2\om_2+t_3\om_3}{ t_1\in \Z,\, t_2,t_3\in \Z^{\geq 0}, -M_1<t_1\leq M_1,\, 3t_2+2t_3\leq M_2  } \cup \\ &\cup  \set{t_1\om_1-t_2\om_2+(t_3+3t_2)\om_3}{ t_1\in \Z,\, t_2,t_3\in \N, -M_1<t_1\leq M_1,\, 3t_2+2t_3< M_2  }.
\end{align*}

The discrete orthogonality relations of the functions $\Xi^{ee}$ are of the form (\ref{diskEE}). We label each point $x=\frac{s_1}{M_1}\om^\vee_1+\frac{s_2}{M_2}\om^\vee_2+\frac{s_3}{M_2}\om^\vee_3\in F^{ee}_{M_1M_2}$ by coordinates $[s_0,s_1,s_0',s_2,s_3]$ such that $s_0+s_1=M_1$ and $s_0'+2s_2+3s_3=M_2$. Similarly, we label each point from $\la=t_1\om_1+t_2\om_2+t_3\om_3\in \Lambda^{ee}_{M_1M_2}$ by coordinates $[t_0,t_1,t_0',t_2,t_3]$ such that $t_0+t_1=M_1$ and $t_0'+3t_2+2t_3=M_2$. The coefficients $\ep^{ee}(x)$ and $h^{ee\vee}_\la$ in (\ref{diskEE}) are listed in Table \ref{TabDiskee}.

\item $W^e(\Aone\times\Gtwo).$ The $W^{e}-$orbit for the generic point $a\om_1+b\om_2+c\om_3\in \R^3 $ is given by
\begin{align*} W^{e}{(a,b,c)}=&\{(a,\pm b,\pm c),(a,\pm
(2b+c),\mp (3b+c)),(a,\mp(b+c),\pm(3b+2c)) \\ &
(-a,\pm b,\mp(3b+c)),(-a,\pm(b+c),\mp c), (-a,\pm(2b+c),\mp(3a+2b))\}\end{align*}
and $|W^{e}|=12$.
The fundamental region $F^{e}$ is given by
\begin{align*}
F^{e}(\Aone\times\Gtwo)=&\{x\omega^\vee_1+y\om^\vee_2+z\om^\vee_3\mid
0\leq x\leq 1,\, y,z\geq 0,\,  2y+3z\leq 1\}\cup\\ & \cup \{x\omega^\vee_1-y\om^\vee_2+(y+z)\om^\vee_3\mid
0< x< 1,\, \, y,z> 0,\,  2y+3z< 1\}
\end{align*}
and its volume is $|F^{e}|=\sqrt{6}/12$.
We have the weight lattice $P_{e}$ of the form
\begin{align*}
P_{e}(\Aone\times\Gtwo)=&\{a\om_1+b\om_2+c\om_3  \mid a, b,c\in \Z^{\geq 0}\} \cup \\ &\cup \{a\om_1-b\om_2+(c+3b)\om_3\mid a,b,c\in \N\}.
\end{align*}
Then for any
$a\omega_1+b\omega_2+c\omega_3\in P_{e}$ the corresponding
$E^{e}-$function at point
$x\omega^\vee_1+y\omega^\vee_2+z\omega^\vee_3$ is
\begin{align*}
\Xi^{e}_{(a,b,c)}(x,y,z) & =2e^{i\pi
ax}(\cos(2\pi((2b+c)y+(3b+2c)z))+\cos(2\pi(ay+(3b+c)z))\\
 &+\cos(2\pi((b+c)y+cz)))+ 2e^{-i\pi
ax}(\cos(2\pi((2b+c)y+(3b+c)z))\\ &+
\cos(2\pi(by-cz))+\cos(\pi((b+c)y+(3b+2c)z)))
\end{align*}
and the coefficients $d^{e}_{(a,b,c)}$ in continuous orthogonality  relations (\ref{e_orthog_e}) have values in Table \ref{TabConte}.

The discrete grid $F^{e}_{M}$ has the explicit form

\begin{align*}
F^{e}_{M}=&\setb{\frac{s_1}{M}\om^\vee_1+\frac{s_2}{M}\om^\vee_2+\frac{s_3}{M}\om^\vee_3}{s_1, s_2,s_3\in \Z^{\geq 0},  s_1\leq M,\, 2s_2+3s_3\leq M   }\cup \\ &\cup \setb{\frac{s_1}{M}\om^\vee_1-\frac{s_2}{M}\om^\vee_2+\frac{s_3+s_2}{M}\om^\vee_3}{s_1, s_2,s_3\in \N,  s_1< M,\, 2s_2+3s_3< M}
\end{align*}
and the corresponding grid $\Lambda^{e}_{M}$ of weights has the following form
\begin{align*}
\Lambda^{e}_{M}=& \set{t_1\om_1+t_2\om_2+t_3\om_3}{ t_1, t_2,t_3\in \Z^{\geq 0}, t_1\leq M,\, 3t_2+2t_3\leq M  } \cup \\ &\cup  \set{t_1\om_1-t_2\om_2+(t_3+3t_2)\om_3}{ t_1, t_2,t_3\in \N,\, t_1< M,\, 3t_2+2t_3< M  }.
\end{align*}

The discrete orthogonality relations of the functions $\Xi^{e}$ are of the form (\ref{orthoEe}). We label each point $x=\frac{s_1}{M}\om^\vee_1+\frac{s_2}{M}\om^\vee_2+\frac{s_3}{M}\om^\vee_3\in F^{e}_{M}$ by coordinates $[s_0,s_1,s_0',s_2,s_3]$ such that $s_0+s_1=M$ and $s_0'+2s_2+3s_3=M$. Similarly, we label each point from $\la=t_1\om_1+t_2\om_2+t_3\om_3\in \Lambda^{e}_{M}$ by coordinates $[t_0,t_1,t_0',t_2,t_3]$ such that $t_0+t_1=M$ and $t_0'+3t_2+2t_3=M$. The coefficients $\ep^{e}(x)$ and $h^{e\vee}_\la$ in (\ref{orthoEe}) are listed in Table \ref{TabDiske}.
\end{enumerate}

\subsection{The $E-$transforms of $\Aone\times\Aone\times\Aone$}\

Relative length and angles between the simple roots and
fundamental weights of $\Aone\times\Aone\times\Aone$ are given in terms
of Cartan matrix
$C=\left(\begin{smallmatrix}2&0&0\\0&2&0\\0&0&2\end{smallmatrix}\right)$, $\det C= 8$.

\begin{enumerate}[1.]
\item $W^{ee}(\Aone\times\Aone\times\Aone)$. The $W^{ee}-$orbit for the generic point $a\om_1+b\om_2+c\om_3\in \R^3 $ is given by
$$ W^{ee}{(a,b,c)}=
\{(a,b, c)\}$$
and $|W^{ee}|=1$.
The fundamental region $F^{ee}$ is given by
\begin{align*}
F^{ee}(\Aone\times\Aone\times\Aone)=&\{x\omega^\vee_1+y\om^\vee_2+z\om^\vee_3\mid
-1< x,y,z\leq 1\}
\end{align*}
and its volume is $| F^{ee}|=2\sqrt{2}$.
We have the weight lattice $P_{ee}$ of the form
\begin{align*}
P_{ee}(\Aone\times\Aone\times\Aone)=&\{a\om_1+b\om_2+c\om_3  \mid a,b,c\in\Z\}
\end{align*}
Then for any
$a\omega_1+b\omega_2+c\omega_3\in P_{ee}$ the corresponding
$E^{ee}-$function at point
$x\omega^\vee_1+y\omega^\vee_2+z\omega^\vee_3$ is
\begin{align*}
\Xi^{ee}_{(a,b,c)}(x,y,z)  = e^{\ii\pi (ax+by+cz)}
\end{align*}
and the coefficients $d^{ee}_{(a,b,c)}$ in continuous orthogonality  relations (\ref{orthog_product1}) have values in Table \ref{TabContee}.

The discrete grid $F^{ee}_{M_1M_2M_3}$ has the explicit form
\begin{align*}
F^{ee}_{M_1M_2M_3}&=\setb{\frac{s_1}{M_1}\om^\vee_1+\frac{s_2}{M_2}\om^\vee_2+\frac{s_3}{M_3}\om^\vee_3}{s_i\in \Z, -M_i<s_i\leq M_i,\,i=1,2,3}
\end{align*}
and the corresponding grid $\Lambda^{ee}_{M_1M_2M_3}$ of weights has the following form
\begin{align*}
\Lambda^{ee}_{M_1M_2M_3}=& \set{t_1\om_1+t_2\om_2+t_3\om_3}{ t_i\in \Z,\,  -M_i<t_i\leq M_i,\, i=1,2,3 }.
\end{align*}

The discrete orthogonality relations of the functions $\Xi^{ee}$ are of the form (\ref{diskEE}) with the resulting normalization coefficient equal to $8M_1M_2M_3h^{ee\vee}_\la $. The coefficients $\ep^{ee}(x)$ and $h^{ee\vee}_\la$ in (\ref{diskEE}) are all equal to $1$.

\item $W^e(\Aone\times\Aone\times\Aone).$ The $W^{e}-$orbit for the generic point $a\om_1+b\om_2+c\om_3\in \R^3 $ is given by
$$ W^{e}{(a,b,c)}=\{(a,\pm b,\pm c),(-a,\pm b,\mp c)\}$$
and $|W^{e}|=4$.
The fundamental region $F^{e}$ is given by
\begin{align*}
F^{e}(\Aone\times\Aone\times\Aone)=&\{x\omega^\vee_1+y\om^\vee_2+z\om^\vee_3\mid
0\leq x,y,z\leq 1 \}\cup\\ & \cup \{-x\omega^\vee_1+y\om^\vee_2+z\om^\vee_3\mid
0< x,y,z< 1\}
\end{align*}
and its volume is $| F^{e}|=1/\sqrt{2}$.
We have the weight lattice $P_{e}$ of the form
\begin{align*}
P_{e}(\Aone\times\Aone\times\Aone)=&\{a\om_1+b\om_2+c\om_3  \mid a, b,c\in \Z^{\geq 0}\} \cup \\ &\cup \{-a\om_1+b\om_2+c\om_3\mid a, b,c\in \N\}.
\end{align*}
Then for any
$a\omega_1+b\omega_2+c\omega_3\in P_{e}$ the corresponding
$E^{e}-$function at point
$x\omega^\vee_1+y\omega^\vee_2+z\omega^\vee_3$ is
\begin{align*}
\Xi^{e}_{(a,b,c)}(x,y,z) & =2e^{\ii\pi ax}\cos\pi(by+cz)+2e^{-\ii\pi
ax}\cos\pi(by-cz)
\end{align*}
and the coefficients $d^{e}_{(a,b,c)}$ in continuous orthogonality  relations (\ref{e_orthog_e}) have values in Table \ref{TabConte}.

The discrete grid $F^{e}_{M}$ has the explicit form

\begin{align*}
F^{e}_{M}=&\setb{\frac{s_1}{M}\om^\vee_1+\frac{s_2}{M}\om^\vee_2+\frac{s_3}{M}\om^\vee_3}{s_1, s_2,s_3\in \Z^{\geq 0},  s_1,s_2,s_3\leq M   }\cup \\ &\cup \setb{-\frac{s_1}{M}\om^\vee_1+\frac{s_2}{M}\om^\vee_2+\frac{s_3}{M}\om^\vee_3}{s_1, s_2,s_3\in \N,  s_1,s_2,s_3< M}
\end{align*}
and the corresponding grid $\Lambda^{e}_{M}$ of weights has the following form
\begin{align*}
\Lambda^{e}_{M}=& \set{t_1\om_1+t_2\om_2+t_3\om_3}{ t_1, t_2,t_3\in \Z^{\geq 0}, t_1,t_2,t_3\leq M  } \cup \\ &\cup  \set{-t_1\om_1+t_2\om_2+t_3\om_3}{ t_1, t_2,t_3\in \N,\, t_1, t_2,t_3< M  }.
\end{align*}

The discrete orthogonality relations of the functions $\Xi^{e}$ are of the form (\ref{orthoEe}) with the resulting normalization coefficient equal to $32M^3h^{e\vee}_\la $. We label each point $x=\frac{s_1}{M}\om^\vee_1+\frac{s_2}{M}\om^\vee_2+\frac{s_3}{M}\om^\vee_3\in F^{e}_{M}$ by coordinates $[s_0,s_1,s_0',s_2,s_0'',s_3]$ such that $s_0+s_1=M$, $s_0'+s_2=M$ and $s_0''+s_3=M$. Similarly, we label each point from $\la=t_1\om_1+t_2\om_2+t_3\om_3\in \Lambda^{e}_{M}$ by coordinates $[t_0,t_1,t_0',t_2,t_0'',t_3]$ such that $t_0+t_1=M$, $t_0'+t_2=M$ and $t_0''+t_3=M$. The coefficients $\ep^{e}(x)$ and $h^{e\vee}_\la$ in (\ref{orthoEe}) are listed in Table~\ref{TabAone3}.
\end{enumerate}

\section{Concluding remarks}\label{Conc}
\begin{itemize}

\item Given a function $f:F^{ee}\map \C$ or $f:F^e\map \C$ we may define interpolating functions in the usual way
\begin{align*}
\Xi^{ee}_{M_1M_2}(x)=&\sum_{\la\in \Lambda^{ee}_{M_1M_2}}\tilde c_{\la}\Xi^{ee}_{\nu}(x),\q x\in \R^{n_1+n_2}\\
\Xi^{e}_{M}(x)=&\sum_{\la\in \Lambda_M^e}c_{\la}\Xi^e_{\la}(x),\q x\in \R^{n_1+n_2}
\end{align*}
where $\tilde c_\la$, $ c_\la$ are given by formulas
(\ref{disc_transform_ee}) and (\ref{disc_transform_e}), respectively. These interpolating functions then coincide with $f$ on the grids $F^{ee}_{M_1M_2}$, $F_M^e$. Interpolation properties of $\Xi^{ee}_{M_1M_2}$, $\Xi^{e}_{M}$ as well as convergence of these functional series deserve further study.
\item Product-to-sum decompositions have for both $E^{ee}-$ and
$E^{e}-$functions the following straightforward form
\begin{align*}
\Xi^{ee}_{\la}(x)\Xi^{ee}_{\la'}(x)=&\sum_{w\in
W^{ee}}\Xi^{ee}_{\la+w\la'}(x),\q x\in \R^{n_1+n_2},\,\la,\la'\in
P_{ee}
\\ \Xi^{e}_{\la}(x)\Xi^{e}_{\la'}(x)=&\sum_{w\in
W^{e}}\Xi^{e}_{\la+w\la'}(x),\q x\in \R^{n_1+n_2},\,\la,\la'\in
P_{e}.
\end{align*}

\item For semisimple $G=G_1\times G_2$ with $W=W_1\times W_2$, there are another possible classes of special functions. One may consider $C-$ or $S-$ functions related to component $W_1$ and $E-$functions related to component $W_2$ or vice versa. Such 'mixed' functions then inherit properties from the corresponding simple components, such as being eigenfunctions of the Laplace operator with different boundary
conditions.
\item Similarly as the common exponential function can be seen as the sum of the cosine and sine functions, the $E-$functions are realized in any dimension as the sum of the appropriate $C-$ and $S-$function. Such functions are considered here for the cases where the underlying Lie group is semisimple but not simple. Recent discovery of additional families of
$W-$invariant(skew-invariant) functions \cite{MotP}, opens several
new possibilities to study of $E$-like functions in our problems.

\end{itemize}

\section*{Acknowledgments}

Work was supported  by the Natural Sciences and Engineering
Research Council of Canada and in part also by the MIND Research
Institute, Santa Ana, California. JH~is grateful for the
postdoctoral fellowship and for the hospitality extended to him at
the Centre de recherches math\'ematiques, Universit\'e de
Montr\'eal. JH also acknowledges partial support by the Ministry
of Education of Czech Republic (project MSM6840770039). IK thanks
to Department of Physics, Czech Technical University for their
hospitality whilst preparing this paper.

\section*{Appendix}
We list the tables of coefficients from section~\ref{Etrans} of
discrete and continuous orthogonality relations of $E^e-$ and
$E^{ee}-$functions. {\small
\begin{table}[ht]
\begin{tabular}{|c|c|}\hline

$\la\in P_e$ & $d^e_\la$  \\ \hline \hline
$(a,b)$ & $1$ \\
$(0,b)$ & $1$ \\
$(a,0)$ & $1$ \\
$(0,0)$ & $2$ \\
 \hline
\end{tabular}\hspace{30pt}
\begin{tabular}{|c|c|}\hline
$x\in F^e_M$ & $\ep^e (x)$  \\ \hline
$[s_0,s_1,s_0',s_2]$ & $2$  \\
$[s_0,s_1,s_0',0]$ & $2$  \\
$[s_0,s_1,0,s_2]$ & $2$  \\
$[s_0,0,s_0',s_2]$ & $2$  \\
$[s_0,0,s_0',0]$ & $1$  \\
$[s_0,0,0,s_2]$ & $1$  \\
$[0,s_1,s_0',s_2]$ & $2$  \\
$[0,s_1,s_0',0]$ & $1$  \\
$[0,s_1,0,s_2]$ & $1$  \\
\hline
\end{tabular}\hspace{24pt}
\begin{tabular}{|c|c|}\hline
$\lambda\in \Lambda^e_M$  & $h_\lambda ^{e\vee}$ \\ \hline
$[t_0,t_1,t_0',t_2]$ & $1$  \\
$[t_0,t_1,t_0',0]$ & $1$  \\
$[t_0,t_1,0,t_2]$ & $1$  \\
$[t_0,0,t_0',t_2]$ & $1$  \\
$[t_0,0,t_0',0]$ & $2$  \\
$[t_0,0,0,t_2]$ & $2$  \\
$[0,t_1,t_0',t_2]$ & $1$  \\
$[0,t_1,t_0',0]$ & $2$  \\
$[0,t_1,0,t_2]$ & $2$  \\  \hline

\end{tabular}
\caption{The coefficients $d_\lambda^{e}$, $\ep^{e}(x)$ and $h^{e\vee}_\la $ of continuous and discrete orthogonality relations of $\Aone\times\Aone$. Assuming $a,b \neq 0$, $s_0,s_1,s_0',s_2\neq 0$ and $t_0,t_1,t_0',t_2 \neq 0$.}\label{TabAone2}
\end{table}

\begin{table}[ht]
\begin{tabular}{|c|c|c|c|c|}\hline\multirow{2}{*}{$\la\in P_{ee}$} & \multicolumn{4}{|c|}{$d^{ee}_\la$} \\
\cline{2-5} & $\Aone\times\Atwo  $ & $\Aone\times\Ctwo  $ & $\Aone\times\Gtwo  $ &$\Aone\times\Aone \times \Aone  $ \\ \hline \hline
$(a,b,c)$ & $1$ & $1$ & $1$ & $1$ \\
$(0,b,c)$ & $1$ & $1$ & $1$ & $1$ \\
$(a,0,c)$ & $1$ & $1$ & $1$ & $1$ \\
$(a,b,0)$ & $1$ & $1$ & $1$ & $1$ \\
$(0,0,c)$ & $1$ & $1$ & $1$ & $1$ \\
$(0,b,0)$ & $1$ & $1$ & $1$ & $1$ \\
$(a,0,0)$ & $3$ & $4$ & $6$ & $1$ \\
$(0,0,0)$ & $3$ & $4$ & $6$ & $1$ \\ \hline
\end{tabular}
\smallskip
\caption{The coefficients $d_\lambda^{ee}$ of continuous orthogonality of semisimple Lie groups of rank $3$.  Assuming $a,b,c \neq 0$. }\label{TabContee}
\end{table}

\begin{table}[ht]
\begin{tabular}{|c|c|c|c|c|}\hline\multirow{2}{*}{$\la\in P_e$} & \multicolumn{4}{|c|}{$d^e_\la$} \\
\cline{2-5} & $\Aone\times\Atwo  $ & $\Aone\times\Ctwo $ & $\Aone\times\Gtwo  $ &$\Aone\times\Aone \times \Aone  $ \\ \hline \hline
$(a,b,c)$ & $1$ & $1$ & $1$ & $1$ \\
$(0,b,c)$ & $1$ & $1$ & $1$ & $1$ \\
$(a,0,c)$ & $1$ & $1$ & $1$ & $1$ \\
$(a,b,0)$ & $1$ & $1$ & $1$ & $1$ \\
$(0,0,c)$ & $2$ & $2$ & $2$ & $2$ \\
$(0,b,0)$ & $2$ & $2$ & $2$ & $2$ \\
$(a,0,0)$ & $3$ & $4$ & $6$ & $2$ \\
$(0,0,0)$ & $6$ & $8$ & $12$ & $4$ \\ \hline
\end{tabular}\bigskip
\caption{The coefficients $d_\lambda^{e}$ of continuous orthogonality of semisimple Lie groups of rank $3$.  Assuming $a,b,c \neq 0$. }\label{TabConte}
\end{table}

\begin{table}[ht]
\begin{tabular}{|c|c|c|c|}\hline
\multirow{2}{*}{$x\in F^{ee}_{M_1M_2}$} & \multicolumn{3}{|c|}{$\ep^{ee}(x)$} \\
\cline{2-4}
& $\Aone\times\Atwo  $ & $\Aone\times\Ctwo $ & $\Aone\times\Gtwo  $  \\ \hline \hline
$[s_0,s_1,s_0',s_2,s_3]$ & $3$ & $4$ & $6$ \\
$[s_0,s_1,s_0',s_2,0]$ & $3$ & $4$ & $6$  \\
$[s_0,s_1,s_0',0,s_3]$ & $3$ & $4$ & $6$  \\
$[s_0,s_1,0,s_2,s_3]$ & $3$ & $4$ & $6$  \\
$[s_0,s_1,s_0',0,0]$ & $1$ & $1$ & $1$  \\
$[s_0,s_1,0,s_2,0]$ & $1$ & $2$ & $2$  \\
$[s_0,s_1,0,0,s_3]$ & $1$ & $1$ & $3$  \\
$[s_0,0,s_0',s_2,s_3]$ & $3$ & $4$ & $6$  \\
$[s_0,0,s_0',s_2,0]$ & $3$ & $4$ & $6$  \\
$[s_0,0,s_0',0,s_3]$ & $3$ & $4$ & $6$  \\
$[s_0,0,0,s_2,s_3]$ & $3$ & $4$ & $6$  \\
$[s_0,0,s_0',0,0]$ & $1$ & $1$ & $1$  \\
$[s_0,0,0,s_2,0]$ & $1$ & $2$ & $2$  \\
$[s_0,0,0,0,s_3]$ & $1$ & $1$ & $3$  \\
$[0,s_1,s_0',s_2,s_3]$ & $3$ & $4$ & $6$  \\
$[0,s_1,s_0',s_2,0]$ & $3$ & $4$ & $6$  \\
$[0,s_1,s_0',0,s_3]$ & $3$ & $4$ & $6$  \\
$[0,s_1,0,s_2,s_3]$ & $3$ & $4$ & $6$  \\
$[0,s_1,s_0',0,0]$ & $1$ & $1$ & $1$  \\
$[0,s_1,0,s_2,0]$ & $1$ & $2$ & $2$  \\
$[0,s_1,0,0,s_3]$ & $1$ & $1$ & $3$  \\
\hline
\end{tabular}  \hspace{5pt}
\begin{tabular}{|c|c|c|c|}\hline
\multirow{2}{*}{$\la\in \Lambda^{ee}_{M_1M_2}$} & \multicolumn{3}{|c|}{$h^{ee\vee}_\la $} \\
\cline{2-4}
& $\Aone\times\Atwo  $ & $\Aone\times\Ctwo $ & $\Aone\times\Gtwo  $  \\ \hline \hline
$[t_0,t_1,t_0',t_2,t_3]$ & $1$ & $1$ & $1$ \\
$[t_0,t_1,t_0',t_2,0]$ & $1$ & $1$ & $1$  \\
$[t_0,t_1,t_0',0,t_3]$ & $1$ & $1$ & $1$  \\
$[t_0,t_1,0,t_2,t_3]$ & $1$ & $1$ & $1$  \\
$[t_0,t_1,t_0',0,0]$ & $3$ & $4$ & $6$  \\
$[t_0,t_1,0,t_2,0]$ & $3$ & $4$ & $3$  \\
$[t_0,t_1,0,0,t_3]$ & $3$ & $2$ & $2$  \\
$[t_0,0,t_0',t_2,t_3]$ & $1$ & $1$ & $1$  \\
$[t_0,0,t_0',t_2,0]$ & $1$ & $1$ & $1$  \\
$[t_0,0,t_0',0,t_3]$ & $1$ & $1$ & $1$  \\
$[t_0,0,0,t_2,t_3]$ & $1$ & $1$ & $1$  \\
$[t_0,0,t_0',0,0]$ & $3$ & $4$ & $6$  \\
$[t_0,0,0,t_2,0]$ & $3$ & $4$ & $3$  \\
$[t_0,0,0,0,t_3]$ & $3$ & $2$ & $2$  \\
$[0,t_1,t_0',t_2,t_3]$ & $1$ & $1$ & $1$  \\
$[0,t_1,t_0',t_2,0]$ & $1$ & $1$ & $1$  \\
$[0,t_1,t_0',0,t_3]$ & $1$ & $1$ & $1$  \\
$[0,t_1,0,t_2,t_3]$ & $1$ & $1$ & $1$  \\
$[0,t_1,t_0',0,0]$ & $3$ & $4$ & $6$  \\
$[0,t_1,0,t_2,0]$ & $3$ & $4$ & $3$  \\
$[0,t_1,0,0,t_3]$ & $3$ & $2$ & $2$  \\
\hline
\end{tabular}
\bigskip
\caption{The coefficients $\ep^{ee}(x)$ and $h^{ee\vee}_\la $ of discrete orthogonality relations of semisimple Lie groups of rank $3$.  Assuming $s_0,s_1,s_0',s_2,s_3\neq 0$ and $t_0,t_1,t_0',t_2,t_3 \neq 0$. }\label{TabDiskee}
\end{table}

\begin{table}[ht]
\begin{tabular}{|c|c|c|c|}\hline
\multirow{2}{*}{$x\in F^e_M$} & \multicolumn{3}{|c|}{$\ep^e(x)$} \\
\cline{2-4}
& $\Aone\times\Atwo  $ & $\Aone\times\Ctwo $ & $\Aone\times\Gtwo  $  \\ \hline \hline
$[s_0,s_1,s_0',s_2,s_3]$ & $6$ & $8$ & $12$ \\
$[s_0,s_1,s_0',s_2,0]$ & $6$ & $8$ & $12$  \\
$[s_0,s_1,s_0',0,s_3]$ & $6$ & $8$ & $12$  \\
$[s_0,s_1,0,s_2,s_3]$ & $6$ & $8$ & $12$  \\
$[s_0,s_1,s_0',0,0]$ & $2$ & $2$ & $2$  \\
$[s_0,s_1,0,s_2,0]$ & $2$ & $4$ & $4$  \\
$[s_0,s_1,0,0,s_3]$ & $2$ & $2$ & $6$  \\
$[s_0,0,s_0',s_2,s_3]$ & $6$ & $8$ & $12$  \\
$[s_0,0,s_0',s_2,0]$ & $3$ & $4$ & $6$  \\
$[s_0,0,s_0',0,s_3]$ & $3$ & $4$ & $6$  \\
$[s_0,0,0,s_2,s_3]$ & $3$ & $4$ & $6$  \\
$[s_0,0,s_0',0,0]$ & $1$ & $1$ & $1$  \\
$[s_0,0,0,s_2,0]$ & $1$ & $2$ & $2$  \\
$[s_0,0,0,0,s_3]$ & $1$ & $1$ & $3$  \\
$[0,s_1,s_0',s_2,s_3]$ & $6$ & $8$ & $12$  \\
$[0,s_1,s_0',s_2,0]$ & $3$ & $4$ & $6$  \\
$[0,s_1,s_0',0,s_3]$ & $3$ & $4$ & $6$  \\
$[0,s_1,0,s_2,s_3]$ & $3$ & $4$ & $6$  \\
$[0,s_1,s_0',0,0]$ & $1$ & $1$ & $1$  \\
$[0,s_1,0,s_2,0]$ & $1$ & $2$ & $2$  \\
$[0,s_1,0,0,s_3]$ & $1$ & $1$ & $3$  \\
\hline
\end{tabular}  \hspace{5pt}
\begin{tabular}{|c|c|c|c|}\hline
\multirow{2}{*}{$\la\in \Lambda^e_M$} & \multicolumn{3}{|c|}{$h^{e\vee}_\la $} \\
\cline{2-4}
& $\Aone\times\Atwo  $ & $\Aone\times\Ctwo $ & $\Aone\times\Gtwo  $  \\ \hline \hline
$[t_0,t_1,t_0',t_2,t_3]$ & $1$ & $1$ & $1$ \\
$[t_0,t_1,t_0',t_2,0]$ & $1$ & $1$ & $1$  \\
$[t_0,t_1,t_0',0,t_3]$ & $1$ & $1$ & $1$  \\
$[t_0,t_1,0,t_2,t_3]$ & $1$ & $1$ & $1$  \\
$[t_0,t_1,t_0',0,0]$ & $3$ & $4$ & $6$  \\
$[t_0,t_1,0,t_2,0]$ & $3$ & $4$ & $3$  \\
$[t_0,t_1,0,0,t_3]$ & $3$ & $2$ & $2$  \\
$[t_0,0,t_0',t_2,t_3]$ & $1$ & $1$ & $1$  \\
$[t_0,0,t_0',t_2,0]$ & $2$ & $2$ & $2$  \\
$[t_0,0,t_0',0,t_3]$ & $2$ & $2$ & $2$  \\
$[t_0,0,0,t_2,t_3]$ & $2$ & $2$ & $2$  \\
$[t_0,0,t_0',0,0]$ & $6$ & $8$ & $12$  \\
$[t_0,0,0,t_2,0]$ & $6$ & $8$ & $6$  \\
$[t_0,0,0,0,t_3]$ & $6$ & $4$ & $4$  \\
$[0,t_1,t_0',t_2,t_3]$ & $1$ & $1$ & $1$  \\
$[0,t_1,t_0',t_2,0]$ & $2$ & $2$ & $2$  \\
$[0,t_1,t_0',0,t_3]$ & $2$ & $2$ & $2$  \\
$[0,t_1,0,t_2,t_3]$ & $2$ & $2$ & $2$  \\
$[0,t_1,t_0',0,0]$ & $6$ & $8$ & $12$  \\
$[0,t_1,0,t_2,0]$ & $6$ & $8$ & $6$  \\
$[0,t_1,0,0,t_3]$ & $6$ & $4$ & $4$  \\
\hline
\end{tabular}
\bigskip
\caption{The coefficients $\ep^{e}(x)$ and $h^{e\vee}_\la $ of discrete orthogonality relations of semisimple Lie groups of rank $3$.  Assuming $s_0,s_1,s_0',s_2,s_3\neq 0$ and $t_0,t_1,t_0',t_2,t_3 \neq 0$. }\label{TabDiske}
\end{table}

\begin{table}[ht]
\begin{tabular}{|c|c|}\hline
$x\in F^e_M$ & $\ep^e (x)$  \\ \hline
$[s_0,s_1,s_0',s_2,s_0'',s_3]$ & $4$  \\
$[s_0,s_1,s_0',s_2,s_0'',0]$ & $4$  \\
$[s_0,s_1,s_0',s_2,0,s_3]$ & $4$  \\
$[s_0,s_1,s_0',0,s_0'',s_3]$ & $4$  \\
$[s_0,s_1,s_0',0,s_0'',0]$ & $2$  \\
$[s_0,s_1,s_0',0,0,s_3]$ & $2$  \\
$[s_0,s_1,0,s_2,s_0'',s_3]$ & $2$  \\
$[s_0,s_1,0,s_2,s_0'',0]$ & $2$  \\
$[s_0,s_1,0,s_2,0,s_3]$ & $2$  \\
$[s_0,0,s_0',s_2,s_0'',s_3]$ & $4$  \\
$[s_0,0,s_0',s_2,s_0'',0]$ & $2$  \\
$[s_0,0,s_0',s_2,0,s_3]$ & $2$  \\
$[s_0,0,s_0',0,s_0'',s_3]$ & $2$  \\
$[s_0,0,s_0',0,s_0'',0]$ & $1$  \\
$[s_0,0,s_0',0,0,s_3]$ & $1$  \\
$[s_0,0,0,s_2,s_0'',s_3]$ & $2$  \\
$[s_0,0,0,s_2,s_0'',0]$ & $1$  \\
$[s_0,0,0,s_2,0,s_3]$ & $1$  \\
$[0,s_1,s_0',s_2,s_0'',s_3]$ & $4$  \\
$[0,s_1,s_0',s_2,s_0'',0]$ & $2$  \\
$[0,s_1,s_0',s_2,0,s_3]$ & $2$  \\
$[0,s_1,s_0',0,s_0'',s_3]$ & $2$  \\
$[0,s_1,s_0',0,s_0'',0]$ & $1$  \\
$[0,s_1,s_0',0,0,s_3]$ & $1$  \\
$[0,s_1,0,s_2,s_0'',s_3]$ & $2$  \\
$[0,s_1,0,s_2,s_0'',0]$ & $1$  \\
$[0,s_1,0,s_2,0,s_3]$ & $1$  \\ \hline
\end{tabular}\hspace{24pt}
\begin{tabular}{|c|c|}\hline
$\lambda\in \Lambda^e_M$  & $h_\lambda ^{e\vee}$ \\ \hline
$[t_0,t_1,t_0',t_2,t_0'',t_3]$ & $1$  \\
$[t_0,t_1,t_0',t_2,t_0'',0]$ & $1$  \\
$[t_0,t_1,t_0',t_2,0,t_3]$ & $1$  \\
$[t_0,t_1,t_0',0,t_0'',t_3]$ & $1$  \\
$[t_0,t_1,t_0',0,t_0'',0]$ & $2$  \\
$[t_0,t_1,t_0',0,0,t_3]$ & $2$  \\
$[t_0,t_1,0,t_2,t_0'',t_3]$ & $2$  \\
$[t_0,t_1,0,t_2,t_0'',0]$ & $2$  \\
$[t_0,t_1,0,t_2,0,t_3]$ & $2$  \\
$[t_0,0,t_0',t_2,t_0'',t_3]$ & $1$  \\
$[t_0,0,t_0',t_2,t_0'',0]$ & $2$  \\
$[t_0,0,t_0',t_2,0,t_3]$ & $2$  \\
$[t_0,0,t_0',0,t_0'',t_3]$ & $2$  \\
$[t_0,0,t_0',0,t_0'',0]$ & $4$  \\
$[t_0,0,t_0',0,0,t_3]$ & $4$  \\
$[t_0,0,0,t_2,t_0'',t_3]$ & $2$  \\
$[t_0,0,0,t_2,t_0'',0]$ & $4$  \\
$[t_0,0,0,t_2,0,t_3]$ & $4$  \\
$[0,t_1,t_0',t_2,t_0'',t_3]$ & $1$  \\
$[0,t_1,t_0',t_2,t_0'',0]$ & $2$  \\
$[0,t_1,t_0',t_2,0,t_3]$ & $2$  \\
$[0,t_1,t_0',0,t_0'',t_3]$ & $2$  \\
$[0,t_1,t_0',0,t_0'',0]$ & $4$  \\
$[0,t_1,t_0',0,0,t_3]$ & $4$  \\
$[0,t_1,0,t_2,t_0'',t_3]$ & $2$  \\
$[0,t_1,0,t_2,t_0'',0]$ & $4$  \\
$[0,t_1,0,t_2,0,t_3]$ & $4$  \\ \hline
\end{tabular}
\caption{The coefficients $\ep^{e}(x)$ and $h^{e\vee}_\la $ of the discrete orthogonality relations of $\Aone\times\Aone\times\Aone$. Assuming $s_0,s_1,s_0',s_2,s_0'',s_3\neq 0$ and $t_0,t_1,t_0',t_2,t_0'',t_3 \neq 0$.}\label{TabAone3}
\end{table}

}

\end{document}

%% file: A1A1ee_Re_01.pdf_t
\begin{picture}(0,0)%
\includegraphics{A1A1ee_Re_01.pdf}%
\end{picture}%
\setlength{\unitlength}{4144sp}%
\begingroup\makeatletter\ifx\SetFigFont\undefined%
\gdef\SetFigFont#1#2#3#4#5{%
  \reset@font\fontsize{#1}{#2pt}%
  \fontfamily{#3}\fontseries{#4}\fontshape{#5}%
  \selectfont}%
\fi\endgroup%
\begin{picture}(1800,1918)(1,-1079)
\put(541,-1006){\makebox(0,0)[lb]{\smash{{\SetFigFont{12}{14.4}{\rmdefault}{\mddefault}{\updefault}{\color[rgb]{0,0,0}{ Re $\Xi^{ee}_{(0, \, 1)}$}}%
}}}}
\end{picture}%

%% file: A1A1ee_Re_10.pdf_t
\begin{picture}(0,0)%
\includegraphics{A1A1ee_Re_10.pdf}%
\end{picture}%
\setlength{\unitlength}{4144sp}%
\begingroup\makeatletter\ifx\SetFigFont\undefined%
\gdef\SetFigFont#1#2#3#4#5{%
  \reset@font\fontsize{#1}{#2pt}%
  \fontfamily{#3}\fontseries{#4}\fontshape{#5}%
  \selectfont}%
\fi\endgroup%
\begin{picture}(1800,1918)(1,-1079)
\put(541,-1006){\makebox(0,0)[lb]{\smash{{\SetFigFont{12}{14.4}{\rmdefault}{\mddefault}{\updefault}{\color[rgb]{0,0,0}{ Re $\Xi^{ee}_{(1, \, 0)}$}}%
}}}}
\end{picture}%

%% file: A1A1ee_Re_m11.pdf_t
\begin{picture}(0,0)%
\includegraphics{A1A1ee_Re_m11.pdf}%
\end{picture}%
\setlength{\unitlength}{4144sp}%
\begingroup\makeatletter\ifx\SetFigFont\undefined%
\gdef\SetFigFont#1#2#3#4#5{%
  \reset@font\fontsize{#1}{#2pt}%
  \fontfamily{#3}\fontseries{#4}\fontshape{#5}%
  \selectfont}%
\fi\endgroup%
\begin{picture}(1800,1918)(1,-1079)
\put(541,-1006){\makebox(0,0)[lb]{\smash{{\SetFigFont{12}{14.4}{\rmdefault}{\mddefault}{\updefault}{\color[rgb]{0,0,0}{ Re $\Xi^{ee}_{(-1, \, 1)}$}}%
}}}}
\end{picture}%

%% file: A1A1ee_Im_01.pdf_t
\begin{picture}(0,0)%
\includegraphics{A1A1ee_Im_01.pdf}%
\end{picture}%
\setlength{\unitlength}{4144sp}%
\begingroup\makeatletter\ifx\SetFigFont\undefined%
\gdef\SetFigFont#1#2#3#4#5{%
  \reset@font\fontsize{#1}{#2pt}%
  \fontfamily{#3}\fontseries{#4}\fontshape{#5}%
  \selectfont}%
\fi\endgroup%
\begin{picture}(1800,1918)(1,-1079)
\put(541,-1006){\makebox(0,0)[lb]{\smash{{\SetFigFont{12}{14.4}{\rmdefault}{\mddefault}{\updefault}{\color[rgb]{0,0,0}{ Im $\Xi^{ee}_{(0, \, 1)}$}}%
}}}}
\end{picture}%

%% file: A1A1ee_Im_10.pdf_t
\begin{picture}(0,0)%
\includegraphics{A1A1ee_Im_10.pdf}%
\end{picture}%
\setlength{\unitlength}{4144sp}%
\begingroup\makeatletter\ifx\SetFigFont\undefined%
\gdef\SetFigFont#1#2#3#4#5{%
  \reset@font\fontsize{#1}{#2pt}%
  \fontfamily{#3}\fontseries{#4}\fontshape{#5}%
  \selectfont}%
\fi\endgroup%
\begin{picture}(1800,1918)(1,-1079)
\put(541,-1006){\makebox(0,0)[lb]{\smash{{\SetFigFont{12}{14.4}{\rmdefault}{\mddefault}{\updefault}{\color[rgb]{0,0,0}{ Im $\Xi^{ee}_{(1, \, 0)}$}}%
}}}}
\end{picture}%

%% file: A1A1ee_Im_m11.pdf_t
\begin{picture}(0,0)%
\includegraphics{A1A1ee_Im_m11.pdf}%
\end{picture}%
\setlength{\unitlength}{4144sp}%
\begingroup\makeatletter\ifx\SetFigFont\undefined%
\gdef\SetFigFont#1#2#3#4#5{%
  \reset@font\fontsize{#1}{#2pt}%
  \fontfamily{#3}\fontseries{#4}\fontshape{#5}%
  \selectfont}%
\fi\endgroup%
\begin{picture}(1800,1918)(1,-1079)
\put(541,-1006){\makebox(0,0)[lb]{\smash{{\SetFigFont{12}{14.4}{\rmdefault}{\mddefault}{\updefault}{\color[rgb]{0,0,0}{ Im $\Xi^{ee}_{(-1, \, 1)}$}}%
}}}}
\end{picture}%

%% file: A1A1e_Re_00.pdf_t
\begin{picture}(0,0)%
\includegraphics{A1A1e_Re_00.pdf}%
\end{picture}%
\setlength{\unitlength}{4144sp}%
\begingroup\makeatletter\ifx\SetFigFont\undefined%
\gdef\SetFigFont#1#2#3#4#5{%
  \reset@font\fontsize{#1}{#2pt}%
  \fontfamily{#3}\fontseries{#4}\fontshape{#5}%
  \selectfont}%
\fi\endgroup%
\begin{picture}(1834,1800)(-11,-961)
\put(766,-781){\makebox(0,0)[lb]{\smash{{\SetFigFont{12}{14.4}{\rmdefault}{\mddefault}{\updefault}{\color[rgb]{0,0,0}{ $\Xi^e_{(0, \, 0)}$}}%
}}}}
\end{picture}%

%% file: A1A1e_Re_01.pdf_t
\begin{picture}(0,0)%
\includegraphics{A1A1e_Re_01.pdf}%
\end{picture}%
\setlength{\unitlength}{4144sp}%
\begingroup\makeatletter\ifx\SetFigFont\undefined%
\gdef\SetFigFont#1#2#3#4#5{%
  \reset@font\fontsize{#1}{#2pt}%
  \fontfamily{#3}\fontseries{#4}\fontshape{#5}%
  \selectfont}%
\fi\endgroup%
\begin{picture}(1834,1800)(-11,-961)
\put(766,-781){\makebox(0,0)[lb]{\smash{{\SetFigFont{12}{14.4}{\rmdefault}{\mddefault}{\updefault}{\color[rgb]{0,0,0}{ $\Xi^e_{(0, \, 1)}$}}%
}}}}
\end{picture}%

%% file: A1A1e_Re_10.pdf_t
\begin{picture}(0,0)%
\includegraphics{A1A1e_Re_10.pdf}%
\end{picture}%
\setlength{\unitlength}{4144sp}%
\begingroup\makeatletter\ifx\SetFigFont\undefined%
\gdef\SetFigFont#1#2#3#4#5{%
  \reset@font\fontsize{#1}{#2pt}%
  \fontfamily{#3}\fontseries{#4}\fontshape{#5}%
  \selectfont}%
\fi\endgroup%
\begin{picture}(1834,1800)(-11,-961)
\put(766,-781){\makebox(0,0)[lb]{\smash{{\SetFigFont{12}{14.4}{\rmdefault}{\mddefault}{\updefault}{\color[rgb]{0,0,0}{ $\Xi^e_{(1, \, 0)}$}}%
}}}}
\end{picture}%

%% file: A1A1e_Re_11.pdf_t
\begin{picture}(0,0)%
\includegraphics{A1A1e_Re_11.pdf}%
\end{picture}%
\setlength{\unitlength}{4144sp}%
\begingroup\makeatletter\ifx\SetFigFont\undefined%
\gdef\SetFigFont#1#2#3#4#5{%
  \reset@font\fontsize{#1}{#2pt}%
  \fontfamily{#3}\fontseries{#4}\fontshape{#5}%
  \selectfont}%
\fi\endgroup%
\begin{picture}(1834,1800)(-11,-961)
\put(766,-781){\makebox(0,0)[lb]{\smash{{\SetFigFont{12}{14.4}{\rmdefault}{\mddefault}{\updefault}{\color[rgb]{0,0,0}{ $\Xi^e_{(1, \, 1)}$}}%
}}}}
\end{picture}%

%% file: A1A1e_Re_m11.pdf_t
\begin{picture}(0,0)%
\includegraphics{A1A1e_Re_m11.pdf}%
\end{picture}%
\setlength{\unitlength}{4144sp}%
\begingroup\makeatletter\ifx\SetFigFont\undefined%
\gdef\SetFigFont#1#2#3#4#5{%
  \reset@font\fontsize{#1}{#2pt}%
  \fontfamily{#3}\fontseries{#4}\fontshape{#5}%
  \selectfont}%
\fi\endgroup%
\begin{picture}(1834,1800)(-11,-961)
\put(766,-781){\makebox(0,0)[lb]{\smash{{\SetFigFont{12}{14.4}{\rmdefault}{\mddefault}{\updefault}{\color[rgb]{0,0,0}{ $\Xi^e_{(-1, \, 1)}$}}%
}}}}
\end{picture}%

%% file: A1A1e_Re_23.pdf_t
\begin{picture}(0,0)%
\includegraphics{A1A1e_Re_23.pdf}%
\end{picture}%
\setlength{\unitlength}{4144sp}%
\begingroup\makeatletter\ifx\SetFigFont\undefined%
\gdef\SetFigFont#1#2#3#4#5{%
  \reset@font\fontsize{#1}{#2pt}%
  \fontfamily{#3}\fontseries{#4}\fontshape{#5}%
  \selectfont}%
\fi\endgroup%
\begin{picture}(1834,1800)(-11,-961)
\put(766,-781){\makebox(0,0)[lb]{\smash{{\SetFigFont{12}{14.4}{\rmdefault}{\mddefault}{\updefault}{\color[rgb]{0,0,0}{ $\Xi^e_{(2, \, 3)}$}}%
}}}}
\end{picture}%

%% file: SemiEfarxiv.bbl
\vspace*{20pt}

\begin{thebibliography}{9}
\bibitem{HP1}
J. Hrivn\'ak, J. Patera, {\it On discretization of tori of compact
simple Lie groups,} J. Phys. A: Math. Theor.~{\bf 42} (2009)
385208; arXiv:0905.2395.

\bibitem{HP2}
J. Hrivn\'ak, J. Patera, {\it On $E-$discretization of tori of
compact simple Lie groups,} J. Phys. A: Math. Theor.~{\bf 43}
(2010) 165206.


\bibitem{Hum}
J.~E.~Humphreys, {\it Introduction to Lie Algebras and
Representation Theory,\/} New York, Springer, 1972.


\bibitem{Kane}
R.~Kane, {\it Reflection Groups and Invariant Theory,\/} New York,
Springer, 2001.

\bibitem{PK}
I.~Kashuba, J.~Patera, {\it Discrete and continuous exponential
transforms of simple Lie groups of rank two,\/} J. Phys A (2007),
{\bf 40} 4751, 23 pages.

\bibitem{KP1}
A. Klimyk, J. Patera, {\it $E$-orbit functions,\/} SIGMA
(Symmetry, Integrability and Geometry: Methods and Applications)
{\bf 4} (2008), 002, 57 pages; arXiv:0801.0822.

\bibitem{MP1}
R.V.~Moody, J.~Patera, {\it Orthogonality within the families of \
$C$-, $S$-, and $E$-functions of any compact semisimple Lie
group,\/} SIGMA (Symmetry, Integrability and Geometry: Methods and
Applications) {\bf 2} (2006) 076, 14 pages, math-ph/0611020.


\bibitem{MotP}
L. Motlochov\'a, J. Patera, {\it Four families of orthogonal polynomials of $C_2$ and symmetric and antisymmetric generalizations of sine and cosine functions}; arXiv:1101.3597

\end{thebibliography}
